\documentclass[floats,floatfix,showpacs,amssymb,prd,twocolumn,superscriptaddress,nofootinbib]{revtex4-1}

\usepackage{amssymb,amsmath,verbatim,mathtools,needspace,enumitem,etoolbox,graphicx,physics,microtype,afterpage,bm}

\usepackage[dvipsnames, usenames]{xcolor}

\definecolor{linkcolor}{rgb}{0.0,0.3,0.5}
\usepackage[unicode, colorlinks=true, linkcolor=linkcolor, citecolor=linkcolor, filecolor=linkcolor,urlcolor=linkcolor, pdfusetitle]{hyperref}
\usepackage[all]{hypcap}
\usepackage[T1]{fontenc}
\usepackage[utf8]{inputenc}
\usepackage{tabularx}
\usepackage{float}
\usepackage{graphicx}
\allowdisplaybreaks
\graphicspath{{./figure/}}

\newcommand{\cuhk}{\affiliation{Department of Physics, The Chinese University of Hong Kong, Shatin, NT, Hong Kong}}
\newcommand{\flatiron}{\affiliation{Center for Computational Astrophysics, Flatiron Institute, New York, NY 10010, USA}}
\newcommand{\CMU}{\affiliation{Physics Department, Carnegie Mellon University, Pittsburgh PA 15213, USA}}
\newcommand{\Princeton}{\affiliation{Department of Astrophysical Sciences, Princeton University, Princeton, NJ 08540, USA}}
\newcommand{\NYU}{\affiliation{Department of Physics, New York university, New York}}

\newcommand{\KU}{\affiliation{Institute for Theoretical Physics, KU Leuven, Celestijnenlaan 200D, B-3001 Leuven, Belgium}}
\newcommand{\KUEE}{\affiliation{Department of Electrical Engineering (ESAT), KU Leuven, Kasteelpark Arenberg 10, B-3001 Leuven, Belgium}}

\definecolor{rb4}{HTML}{27408B}

\newcommand\myeq{\stackrel{\mathclap{\footnotesize	\mbox{PCA}}}{\Rightarrow}}

\begin{document}

\title{Testing the robustness of simulation-based gravitational-wave population inference}

\pacs{}

\author{Damon H. T. Cheung} 
\email{damoncheung@link.cuhk.edu.hk}
\cuhk
\author{Kaze W. K. Wong} 
\flatiron

\author{Otto A. Hannuksela} 
\cuhk 
%\nikhef
%\GRASP

\author{Tjonnie G. F. Li} 
\cuhk
\KU 
\KUEE

\author{Shirley Ho}
\flatiron
\Princeton
\CMU
\NYU

\date{\today}
\begin{abstract}
Gravitational-wave population studies have become more important in gravitational-wave astronomy because of the rapid growth of the observed catalog.
In recent studies, emulators based on different machine learning techniques are used to emulate the outcomes of the population synthesis simulation quickly. 
In this study, we benchmark the performance of two emulators that learn the truncated power law phenomenological model by using Gaussian process regression and normalizing flows technique to see which one is a more capable likelihood emulator in the population inference. 
We benchmark the characteristic of the emulators by comparing their performance in the population inference to the phenomenological model using mock and real observation data.
Our results suggest that the normalizing flows emulator can recover the posterior distribution by using the phenomenological model in the population inference with up to 300 mock injections.
The normalizing flows emulator also underestimates the uncertainty for some posterior distributions in the population inference on real observation data.
On the other hand, the Gaussian process regression emulator has poor performance on the same task and can only be used effectively in low-dimension cases.

\end{abstract}

\maketitle

\section{Introduction} \label{sec:intro}
%Since the first direct observation of gravitational-wave (GW) event \cite{firstGW} was announced by Laser Interferometer Gravitational-Wave Observatory (LIGO)\cite{LIGO} in 2015, GW events are being detected routinely at a rapid pace. 
Since the first detection of a gravitational-wave (GW) event was announced by LIGO-Virgo Collaboration in 2016 \cite{firstGW}, GW events are being detected routinely at a rapid pace.
%Since the first direct observation of gravitational-wave (GW) event was announced by Laser Interferometer Gravitational-Wave Observatory (LIGO) and Virgo in 2015 \cite{firstGW}, GW events are being detected routinely at a rapid pace. 
Construction of second-generation global GW detection networks was done in the past five years including the upgrade of Advanced LIGO/Virgo \cite{AdvancedLIGO,AdvancedVIRGO} and the Kamioka Gravitational Wave Detector (KAGRA) in Japan \cite{KAGRA}. 
We are getting more observed GW events from these detectors and recently, the LIGO-Virgo-KAGRA Collaboration (LVK) has announced 35 candidate events in the second half of the third observational run (O3b) \cite{GWTC3}.
By adding the observed events in the first, second, and the first half of the third observational run (O1, O2, and O3a) \cite{GWTC1,GWTC2}, the third Gravitational-Wave Transient Catalog (GWTC-3) contains over 60 events across the first three observing runs \cite{popgwtc3}.
Moreover, there is proposed construction of more powerful detectors with higher sensitivity, such as Einstein Telescope in Europe \cite{EinsteinTelescope} and Cosmic Explorer in the USA \cite{CosmicExplorer}. 
The rapid growth of observed events in the catalog is forecasted as $\sim 10^6$ GW events to be observed per year in this third-generation detectors network \cite{prospects_LVK}. 
It opens up a unique window to study the population properties of compact objects [e.g., black holes (BHs) and neutron stars (NSs)] which helps to find the fundamental physics of the Universe.
For example, the population studies can improve the existing GW constraints on theory-agnostic modifications to general relativity and explore gravity theories beyond general relativity \cite{prob_population_fundamental_physics}.

Typical analyses assume the underlying population of GW sources is described by some phenomenological model (e.g., \cite{PhysRevD.97.043014},\cite{PhysRevD.100.043012}). For instance, one can assume the mass distribution of a merging binary BH follows a power law. Such a simple assumption of the likelihood can provide a low computational cost and a clear statistical interpretation behind it. 
As the size of the GW catalog grows rapidly, a more complex phenomenological model is needed to capture a more sophisticated statistical relation. 
However, phenomenological models do not come from a first principle physics simulation or calculation.
Alternatively, one can use population synthesis simulations that are based on some physical parameters such as the BH natal kick velocity \cite{stellarmass_natalkick}, the common envelope efficiency of binary evolution \cite{m1}, and the metallicity of the environment \cite{metallicity}.
The population synthesis simulations simulate complex physics rather than only describing the structure of the distribution.
However, they take a long time to simulate and have high computational costs.  
In this case, we can use a nonparametric density estimator to emulate the outcomes of the population synthesis simulation with fast speed.
Then, we can perform the population inference efficiently by using the estimator to provide direct physical insights into GW population studies.

There are recent developments on emulating simulations through machine learning to build the population probability density emulator for GW population studies \cite{taylor2018mining,PhysRevD.100.083015,Wong:2020jdt}. 
The emulator interpolates simulation output without going through sophisticated simulations that are fast enough to be used in hierarchical Bayesian analysis (HBA) for population inference.
One of these techniques is to combine Gaussian process regression (GPR), principal component analysis (PCA), and space-filling algorithms to train the emulator \cite{taylor2018mining}.
Another example is applying a deep generative flow technique; normalizing flows (NF) \cite{Wong:2020jdt} to train the emulator. 
However, we lack benchmark results to show how well they perform and what limitations they have when utilized in the GW population inference.

In this study, we investigate the performance of the emulators trained by using these two techniques, which we refer to as the GPR emulator and the NF emulator.
We train the emulators to learn the truncated power law phenomenological model \cite{POPgwtc2} and demonstrate their ability by comparing the performance in event sampling and population inference to the phenomenological model.
More specifically, we use the emulators to sample GW events to see if the distributions match the phenomenological model.
We also implement the emulators in the HBA framework to act as the population probability density emulator and compare the sampled posterior distribution to the phenomenological model by injecting mock and GWTC-2 data. 
Note that the selection bias caused by the limitation of instruments \cite{instrubias} will bias the sampled posterior distribution in the population inference \cite{PhysRevD.100.083015,selectionbias_pop_inference}.
Therefore, we account for the selection bias in the population inference when benchmarking the emulators.

This paper is structured as follows. 
In Sec. \ref{sec:method}, we review the GW population data analysis pipeline.
In Sec. \ref{emulator}, we describe the details of the two machine learning emulators we used in this paper. 
In Sec. \ref{result}, we present the emulators' performance by using both mock data and GWTC-2 data. 
And lastly, in Sec. \ref{sec:discussion}, we discuss the limitation of the emulators and the future directions of this work.

\section{Method} \label{sec:method}

%\kw{Start with a more general introduction to HBA.}
\subsection{Hierarchical Bayesian analysis} \label{sec:HBA}
In this section, we summarize the pipeline of GW population data analysis. We start with the data analysis on a single GW event. Then we show how we make use of a set of observed GW events to study the population properties using the HBA.
First, GW data $\bm{d}$ announced by LVK is usually in the form of a time series which contains no physical quantities directly. 
The data can be modeled using a waveform model characterized by source properties $\bm{\theta}$ (e.g., masses, masses ratio, redshift, spin) known as the event parameters.
In order to extract physical quantities from a time series, we often use Bayesian inference to adopt a parameter estimation process \cite{para_estimation}. 
Given a time series $\bm{d}$, the event posterior $p(\bm{\theta}|\bm{d})$ can be obtained using Bayes' theorem, 
\begin{equation}
\label{eq:event_bayes}
p(\bm{\theta}|\bm{d}) = \pi(\bm{\theta})\frac{p(\bm{d}|\bm{\theta})}{p(\bm{d})},
\end{equation}
where $\pi(\bm{\theta})$ is the prior of event parameters, $p(\bm{d}|\bm{\theta})$ is the event likelihood of observing the data given the source properties with a specific waveform model, and $p(\bm{d})$ is the evidence. $\pi(\bm{\theta})$ carries the physical intuition (e.g., mass cannot be negative and masses ratio cannot be greater than 1) which can also affect the estimation result \cite{PhysRevLett.119.251103,prior}.
%With Bayesian inference, we sample the event posterior of a GW event using the Markov Chain Monte Carlo (MCMC) which can be performed using \textbf{emcee} \cite{foreman2013emcee}. 

To study the population, we can employ a phenomenological population model or simulation-based model which is characterized by the hyperparameters $\bm{\lambda}$ and then infer the hyperparameters favored by the observed catalog \cite{DavidHogg_HBA}.
For example, if we take the route of employing a phenomenological model, the distribution of mass can follow a power law with spectral index $\alpha$, i.e., $p(\bm{\theta}|\bm{\lambda}) = p(m|\alpha)\sim m^{\alpha} $ with $\alpha$ being the hyper-parameter. 
On the other hand, we can employ a simulation-based model that can provide a synthetic catalog of GW events instead of an analytical expression of the population probability density function. 
In this case, we can train an emulator on this model using machine learning techniques to emulate the $p(\bm{\theta}|\bm{\lambda})$, where the hyperparameters could be some physical parameters such as the metallicity of the environment \cite{metallicity}.
In the following, we summarize a statistical framework of inferring the hyperparameters given a set of observations in the content of the GW population. 
The framework is commonly labeled as hierarchical modeling \cite{DavidHogg_HBA}. 
We refer interested readers to more detailed explanations in the literatures \cite{DavidHogg_HBA,stat_frame_sb,GWHBA}. 
Similar to the parameter estimation of a single GW event, we now want to infer the hyperparameters of the population model given some time series data. Therefore, we start by writing down Bayes' theorem in terms of $\bm{d},$ and $\bm{\lambda}$ as
\begin{equation}
\label{eq:bayes}
p(\bm{\lambda}|\bm{d}) = \pi(\bm{\lambda})\frac{p(\bm{d}|\bm{\lambda})}{p(\bm{d})},
\end{equation}
where $p(\bm{\lambda}|\bm{d})$ is the population posterior, $p(\bm{d}|\bm{\lambda})$ is the likelihood of observing the data set given the population model characterized by the hyperparameters $\bm{\lambda}$, $\pi(\bm{\lambda})$ is the prior of hyperparameters, and $p(\bm{d})$ is the evidence.
However, population synthesis simulations give the population in terms of event parameters instead of time series. Therefore, we need to expand the marginalized likelihood $p(\bm{d}|\bm{\lambda})$ as $p(\bm{d}|\bm{\lambda}) = \int p(\bm{d}|\bm{\theta}) p_{\mathrm{pop}}(\bm{\theta}|\bm{\lambda})d\bm{\theta}$ and replace $p(\bm{d}|\bm{\theta})$ by using Eq. [\ref{eq:event_bayes}] to get
\begin{equation}
\label{eq:bayes_event}
p(\bm{\lambda}|\bm{d}) = \pi(\bm{\lambda})\int \frac{ p(\bm{\theta}|\bm{d}) p_{\mathrm{pop}}(\bm{\theta}|\bm{\lambda}) }{\pi(\bm{\theta})}d\bm{\theta},
\end{equation}
where $ p_{\mathrm{pop}}(\bm{\theta}|\bm{\lambda})$ is the population probability density of observing the event given the population model characterized by the hyperparameters. 

Furthermore, if the dataset contains multiple observed GW events that are drawn independently from the population, in other words, the signals are not overlapping and the parameter
estimation is not correlated for different events, we can separate the likelihood of observing that particular set of events [i.e., $p(\bm{d}|\bm{\lambda})$ in Eq. [\ref{eq:bayes}]], the integral in Eq. [\ref{eq:bayes_event}] into the product of the individual likelihoods. 
Therefore, we can rewrite Eq. [\ref{eq:bayes_event}] as
\begin{equation}
\label{eq:product_hba}
p(\bm{\lambda}|\bm{d}) = \pi(\bm{\lambda}) \prod_{i=1}^{N_{\mathrm{obs}}} \int \frac{p_i(\bm{\theta}_i|\bm{d}_i)}{\pi_i(\bm{\theta}_i)} p_{\mathrm{pop}}(\bm{\theta}_i|\bm{\lambda})d\bm{\theta},
\end{equation}
where $\bm{d}_i$ refers to the segment of the whole time series which contains the $i$th event characterized by $\bm{\theta}_i$ and $N_{\mathrm{obs}}$ is the number of observed events.
By separating the likelihood into a product of individual likelihoods, we assume the event parameter estimation is not correlated. 
This is a valid assumption for the current generation detector. 
However, we need to revise the assumption for detectors in the next generation, such as the Einstein Telescope and Cosmic Explorer.
They can detect a large number of GW signals and may eventually overlap, and the interference of the overlapped waveform may affect the parameter estimation and thus the population inference \cite{3rdoverlap_signal}.

\begin{figure*}[htb!]
    \centering
    \includegraphics[width=0.8\textwidth]{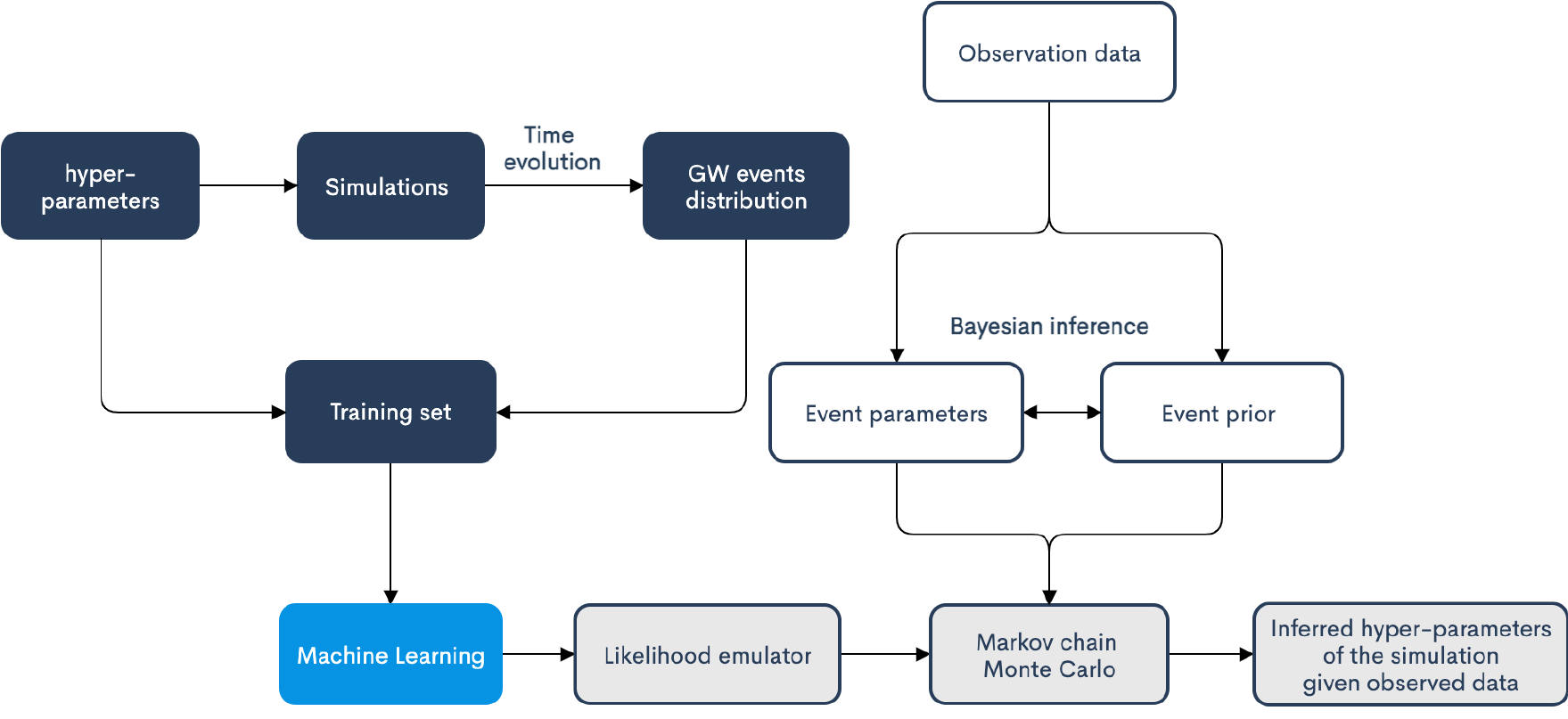}
    \caption{Schematic diagram of GW event population data analysis pipeline. Training data is generated by running a set of simulations with different hyperparameters inputs. Then we use the training data to train the likelihood emulator with a machine learning technique. Event parameters and priors are obtained by performing Bayesian inference on observation data. Lastly, given the event parameters, prior and the population probability density emulator, we sample the posterior of hyperparameters using the MCMC method.}
    \label{fig:HBAframework} 
\end{figure*}

In real life data, there is often a selection bias that comes from the limitations of ground-based interferometers. 
The detectors can only detect signals from a specific frequency range above a signal-to-noise ratio threshold \cite{sensitivity_ligovirgo}. 
It limits the ability to detect weak signals and misses many low mass GW events.
In addition, the sensitivity of the detectors depends on the sky location \cite{instrubias} and the luminosity distance correlates with the redshift of the event \cite{sensitivity_1,sensitivity_2}.
As a result, some events are easier to observe than others, which introduces a bias on the observed population.
Whether we can detect the event is not a binary yes or no because the detectors are noisy.
The best we can do is to calculate $p_{\mathrm{det}}(\bm{\theta})$ i.e., the probability of detecting an event with event parameters $\bm{\theta}$ \cite{selectionbias_pop_inference}.
When we account the selection bias, Eq. [\ref{eq:product_hba}] becomes    
\begin{equation}
\label{eq:int_selectionbias}
p(\bm{\lambda}|\bm{d}) = \pi(\bm{\lambda}) \prod_{i=1}^{N_{\mathrm{obs}}} \int \frac{p_i(\bm{\theta}_i|\bm{d}_i)}{\pi_i(\bm{\theta}_i)} 
\frac{p_{\mathrm{pop}}(\bm{\theta}_i|\bm{\lambda})}{\alpha(\bm{\lambda})}d\bm{\theta},
\end{equation}
where $\alpha(\bm{\lambda})=\int p_{\mathrm{pop}}(\bm{\theta}'|\bm{\lambda})p_{\mathrm{det}}(\bm{\theta}') d\bm{\theta}'$ is the selection bias term.
%\kw{Add reasons why we use reweighting} 
Notice that the computation of multidimensional $\bm{\theta}$ integrals are very expensive. 
Therefore, the event posterior $p_i(\bm{\theta}_i|\bm{d}_i)$ is often given in a form of discrete samples from event parameter estimation \cite{bilby_para_est,PhysRevD.91.042003}.
We can then separate the event parameter estimation from sampling the population posterior.
First, we perform event parameter estimation and save the event posterior samples. 
Then, we compute $p(\bm{\lambda}|\bm{d})$ using the event posterior samples to avoid unnecessary recomputation of event parameter estimation which significantly reduces the computation load for each run.
The above process is equivalent to computing the integral in Eq. [\ref{eq:int_selectionbias}] as the expectation value of the population probability density that has been reweighted by the prior. That is, replace the integral with the sum of the discrete-event posterior samples as
\begin{equation}
\label{eq:governing_HBA}
p(\bm{\lambda}|\bm{d}) = \pi(\bm{\lambda}) \prod_{i=1}^{N_{\mathrm{obs}}} \frac{1}{S_i} \sum_{j=1}^{S_i} \frac{p_{\mathrm{pop}}(\bm{\theta}_{i}^{j}|\bm{\lambda})}{\pi(\bm{\theta}_{i}^{j})\alpha(\bm{\lambda})},
\end{equation}
where $j$ labels the $j$-th posterior sample of the $i$th event and $S_i$ is the number of discrete posterior samples for the $i$th event.
Notice that we do not include event rate in our derivation so 
Eq. [\ref{eq:governing_HBA}] is the governing equation for the HBA framework.
Once we get all the ingredients, we can sample the posterior using various methods such as nested sampling and Markov Chain Monte Carlo (MCMC). In this study, we use MCMC to sample the posterior. The samples represent $p(\bm{\lambda}|\bm{d})$ which tells us the inferred hyperparameters of the simulation that favored by the observation data. We summarize the pipeline in a schematic diagram shown in Fig. \ref{fig:HBAframework}.

\subsection{Computation of selection bias}
In order to compute $\alpha(\bm{\lambda})$ in Eq. [\ref{eq:governing_HBA}], one needs to inject a large amount of signals and recover them with a search pipeline to estimate $\alpha(\bm{\lambda})$; this incurs an expensive computational cost not to mention $\alpha(\bm{\lambda})$ will be computed for each step in the MCMC.
Therefore, we approximate it via Monte Carlo with importance sampling \cite{selectionbias}.
By drawing events from a known distribution $\bm{\theta} \sim p_{\mathrm{draw}}$, we can then get the selection bias term by averaging the population probability of detectable events over the drawn samples.
as
\begin{equation}
\label{eq:selectionbias}
    \alpha(\bm{\lambda}) \approx \frac{1}{N_{\mathrm{draw}}} \sum_{j=1}^{N_{\mathrm{det}}} \frac{p_{\mathrm{pop}}(\bm{\theta}|\bm{\lambda})}{p_{\mathrm{draw}}(\bm{\theta})},
\end{equation}
where $j$ labels the $j$th detectable sample of drawn events, $N_{\mathrm{draw}}$ is the number of event samples to be drawn from $p_{\mathrm{draw}}$, and $N_{\mathrm{det}}$ is the number of detected events from drawn event samples.

When we inject the O1 + O2 + O3a catalog into the GW population data analysis pipeline, $\alpha(\bm{\lambda})$ is evaluated by reweighting an injection campaign done by the LVC \cite{POPgwtc2}.
When we test our pipeline by injecting mock data, $\alpha(\bm{\lambda})$ is evaluated by using the $p_{\mathrm{det}}(\bm{\theta})$ function in the developed interpolation package \textbf{gwdet}\cite{davide_gerosa_2017_889966} for simplicity.

\section{Emulator} \label{emulator}
In computing the population posterior $p(\bm{\lambda}|\bm{d})$ in Eq. [\ref{eq:governing_HBA}], the population probability density function $p_{\mathrm{pop}}(\bm{\theta}|\bm{\lambda})$ is the most important part since it relates the event distribution and the hyperparameters of the simulation.
A common approach is writing down an analytic or semianalytic population probability density function.
However, in the case of population synthesis simulation, it is not so simple.
As a result, we need to simulate each sampling step to calculate $p(\bm{\lambda}|\bm{d}))$ in Eq. [\ref{eq:governing_HBA}].
%For instance, a typical MCMC requires 40 walkers to walk 1000 steps in hyper-parameter space that needs $4\times 10^4$ simulations to be done. 
%But running a typical population synthesis simulation needs $\sim 3-4$ hours which is expensive in the population pipeline \cite{popsyn_compactObjectModeling,popsyn_EvolutionOfBinary,popsyn_MOCCA,popsyn_progenitorsOfCompactObject}.
In addition, a typical population synthesis simulation takes $\sim 3-4$ hours to complete.
If we simulate each step, the population pipeline becomes computationally expensive \cite{popsyn_compactObjectModeling,popsyn_EvolutionOfBinary,popsyn_MOCCA,popsyn_progenitorsOfCompactObject}.
Hence, we train a emulator by using machine learning techniques to learn the likelihood function $p_{\mathrm{pop}}(\bm{\theta}|\bm{\lambda})$ from population synthesis simulations. 
A trained emulator can approximate the output of simulation by giving the hyperparameters $\bm{\lambda}$ without going through the sophisticated simulations. 
To benchmark the capability of the emulators, we test whether they can recover the phenomenological model likelihood and compare their performance in the HBA framework. 
% not verify in general just test again simple model (capability)
The inferred posterior by using the phenomenological model is treated as a control set for the comparison. 
In this study, truncated power law phenomenological model \cite{POPgwtc2} (see Appendix \ref{A1}) is chosen for the comparison with the GPR emulator and the NF emulator respectively.
The truncated power law phenomenological model's event parameters are $m_1,m_2$ and $z$ which correspond to the primary mass, secondary mass and redshift of the GW event, respectively, while the hyperparameters $\bm{\lambda} = [ \alpha,\beta,m_{\mathrm{min}},m_{\mathrm{max}} ]$.
In the following three subsections, we will present the method used for generating training data. Then, we review how we train our GPR emulator and NF emulator.  

\subsection{Training data} \label{sec:data}
%\kw{Short enough, just merge with previous paragraph.}
In this study,  we use 400 simulations as our training set. We choose the hyperparameters $\bm{\lambda}_{\mathrm{train}} = \{ \bm{\lambda}_i\},i=1,2,...,400 $ of the simulations by Latin-hypercube sampling (LHS). 
LHS gives the advantage that stratifies each univariate margin simultaneously \cite{lhsestimates} with variance reduction form compared with uniform random sampling \cite{lhs_stein}. 
It does not contain a duplicate number in each hyper-parameter dimension. 
Therefore, LHC gives more uniform coverage in the hyper-parameter space than Cartesian grid sampling to help the training.
We use python package \textbf{pyDOE} \cite{pyDOE} to carry out the LHS.
Then we draw $10^5$ events from each simulation by rejection sampling as the training set for both emulators. 100 more simulations are generated as a validation set for training the NF emulator. 

\subsection{Gaussian process regression emulator} \label{sec:gpr}
%Gaussian process regression is a non-parametric density estimation method that estimates the distribution by inferring the training data with Gaussian prior \cite{gpr}. 
Gaussian process regression is a nonparametric density estimation method.
Instead of parametrizing the underlying density function, it places a Gaussian prior characterized by a mean and covariance to describe the possible density function. Then we can fit the density function by inferring the mean and covariance with the training data \cite{gpr}.
We follow the method in \cite{taylor2018mining} tightly to construct the GPR emulator.
First, we produce histograms with equal-sized bins over event parameters to summarize the event distribution for simulations characterized by different hyperparameters.
Then, we form a matrix $A_{m\times n}$ by using the information of histograms, where $m$ is the number of simulations in the training set and $n$ is the number of flatten bins over all event parameters in the histograms. 
The probability of having the event (represented by the bin) is proportional to the height of each histogram bin; therefore, we can apply GPR to learn how the input hyperparameters affect the height.
However, some components of the basis obtained by such naive binning might be unnecessary if the training data can be described only by some main features.
It will increase the computational cost exponentially.
Therefore, before applying GPR, we use PCA to form a new set of data-driven basis which is smaller in number than the basis obtained in the naive binning method.
%\kw{This paragraph is quite confusing. I guess your idea is to say we start from a full rank matrix, decompose it using SVD then ditch insignificant basis, then rotate back into original simulation basis to reduce data complexity. Need to be written.}
PCA decomposes the data matrix $A_{m\times n}$ as 
\begin{equation}
\label{eq:svd}
A_{m\times n} = U_{m\times m} S_{m\times n} W_{n\times n}^T 
\end{equation}
where $U$ and $W$ are constituted by orthonormal eigenvectors chosen from $AA^T$ and $A^TA$ respectively, $A^T$ is the transpose of $A$. $S_{m\times n}$ is a positive-semidefinite matrix which can be interpreted as a rectangular diagonal matrix with the variance $\sigma_m$ of each basis.
With this form, we can then eliminate the basis corresponding to $\sigma_{m} <\epsilon$ to reduce the dimensions of $U, S, W$ as
\begin{align}
\label{eq:pca}
\nonumber A_{m\times n} &\myeq (A_{m\times n} )_{\sigma_m > \epsilon} \\
                        &=\tilde{A}_{m'\times n'} \\
\nonumber               &=\tilde{U}_{m'\times m'} \tilde{S}_{m'\times n'}  \tilde{W}_{n'\times n'}^T\\
\nonumber               &=(\frac{\tilde{U}_{m'\times m'} \tilde{S}_{m'\times n'}}{\sqrt{N_{\mathrm{basis}}}}) (\sqrt{N_{\mathrm{basis}}} \tilde{W}_{n'\times n'}^T), 
\end{align}
%and get the elements of PCA according to Eq. [\ref{eq:pca}]. 
where $\epsilon$ is a small number, $N_{\mathrm{basis}}$ is the number of basis after reducing dimensions and $\tilde{U},\tilde{S},\tilde{W}$ are formed by restricting $U, S, W$ on a basis with $\sigma_{m} <\epsilon$ condition.
The columns of $\tilde{U}_{m'\times m'} \tilde{S}_{m'\times n'} / \sqrt{N_{\mathrm{basis}}}$ are the principal components (PCs) of the data matrix, while $\sqrt{N_{\mathrm{basis}}} \tilde{W}_{n'\times n'}^T$ is the projection of the original histogram heights into the new basis.
This helps to reduce data complexity without losing too much information as the basis after PCA describes the main features of the whole training data set. 
Then, we use \textbf{scikit-learn} \cite{scikitlearn} to apply GPR on each basis with correct PC weighting by inferring training data with a Gaussian prior.
The trained emulator gives the resulting posterior-predictive distribution with Gaussian noise that comes from the credible region.
We can obtain a point prediction of $p_{\mathrm{pop}}(\bm{\theta}|\bm{\lambda})$ using the mean of the posterior distribution. 

In this study, we are not able to obtain a satisfactory GPR emulator by using 400 simulations, each containing $10^5$ or even $10^6$ events as the training data.
Therefore, we construct the matrix $A_{m\times n}$ using the theoretical probability density of the phenomenological model as the height for each histogram bin.
After reducing the complexity using PCA, we keep 114 PCs to train the emulator.

\subsection{Normalizing flows emulator} \label{sec:nf}
Another approach we use to emulate conditional probabilities is using conditional neural density estimators, in particular, a flow-based generative (often referred to as normalizing flows) model \cite{MAF}.
Unlike other neural density estimators using variational autoencoders \cite{variationalBayes} or generative adversarial networks \cite{adversarial} that can only generate new data that mimics the target distribution (in our case, the GW event distribution of the simulation), a flow-based generative model can also provide an estimate of the probability density which can be evaluated fast enough in HBA. 
In this section we present the basic principles behind the model we use.

The idea of NF is to transform a simple probability density (e.g., a Gaussian ) $z\sim p_z$ into a target probability density which is much more complicated $x \sim p_x$  by an invertible transformation with tractable Jacobian. The transformation is a mapping function $ \bm{g}: \mathbb{R}^d \rightarrow \mathbb{R}^d $ for $\bm{x},\bm{z} \in \mathbb{R}^d$.
% see wiki for their introduction/ explanation (jacob part) 
%The core idea is the change of variables. If we describe a density distribution in $\mathbb{R}^d$ by two different parameterizations $ \bm{x,z} \in \mathbb{R}^d$ with proper normalization, they need to satisfy the condition 
We can then get the change of variable relation from the normalization condition of probabilities as
\begin{align}
\label{eq:change of variable}
\nonumber p_x(\bm{x})&=p_z(\bm{z})  \left| \det \frac{\partial\bm{z}}{\partial \bm{x}}\right| \\
                     &=p_z(\bm{g}^{-1} (x))  \left| \det \frac{\partial\bm{g}^{-1} (x)}{\partial \bm{x}}\right|,
\end{align}
which requires the transformation to be invertible and thus gives a tractable Jacobian to evaluate $p_x(\bm{x})$.
For estimating more complex high-dimensional distribution, we need more complex transformations, which can be done by applying a series of invertible transformations as
\begin{equation}
\label{eq:5}
\bm{x} = \bm{z}_k=\bm{g}_k \circ \bm{g}_{k-1}\circ \cdots \circ \bm{g}_1(z_0),
\end{equation}
where $\bm{z}_k$ is the distribution after the $k$th transformation function. The condition on invertibility is still fulfilled since the composition of invertible functions is invertible.
At the same time, we need to be aware of the time for training since the computation of high-dimensional Jacobian determinants is expensive. In conclusion, the transformations $\bm{g}$ should be invertible and simple.
Then, we can write down the overall transformation as 
\begin{align}
\label{eq:chain of change of variable}
\nonumber p_x(\bm{x}) &= p_{z_0}(\bm{z}_0)\prod_{k=1}^{N_{\mathrm{transf}}} \left| \det \frac{\partial \bm{z}_{k-1}}{\partial \bm{z}_{k}}\right|\\
                      &= p_{z_0}(\bm{z}_0)\prod_{k=1}^{N_{\mathrm{transf}}} \left| \det \frac{\partial \bm{g}^{-1}_k(z_{k})}{\partial \bm{z}_{k}}\right|,
\end{align}
where $N_{\mathrm{transf}}$ is the number of transformations.

However, choosing the correct transformation is crucial to designing an efficient network for a specific problem. Our target is to emulate $p_{\mathrm{pop}}(\bm{\theta}|\bm{\lambda})$ so the network should be capable to model conditional probabilities. 
A specifically designed flow-based generative model known as masked autoregressive flow (MAF) \cite{MAF} is capable of such a purpose. It is a particular implementation of the NF that uses the masked autoencoder for distribution estimation (MADE) \cite{papamakarios2017masked} as a building block instead of the fully-connected layer.
MADE masks some autoencoder’s parameters of hidden layers to respect autoregressive constraints that each node is only from previous inputs in a given ordering so that the node only depends on some nodes from the previous layer. 
It expands the joint probability into the products of the conditional probabilities' relation with a different order \cite{MADE_product}.
Therefore, we use MAF with a 10 layer network as the “flow” in the emulator. 

We can then train the emulator with a loss function defined as
\begin{equation}
\label{eq:loss}
\mathcal{L} = - \frac{1}{\left|\bm{\mathcal{D}}\right|}\sum_{\bm{x}\in \bm{\mathcal{D}}} \log (p_x{(\bm{x}})),
\end{equation}
where $\mathcal{L}$ is the loss function, $\bm{\mathcal{D}}$ is the dataset and $p_x{(\bm{x})}$ is the entire transformation.
$\mathcal{L}$ gives the indicator whether the emulator provides a similar distribution compared to target distribution $\bm{x}$. Then, the emulator with the best transformation is obtained by finding the global minimum of $\mathcal{L}$ \cite{liao2021jacobian}.
\begin{figure*}[t]
    \centering
    \includegraphics[width=0.45\textwidth]{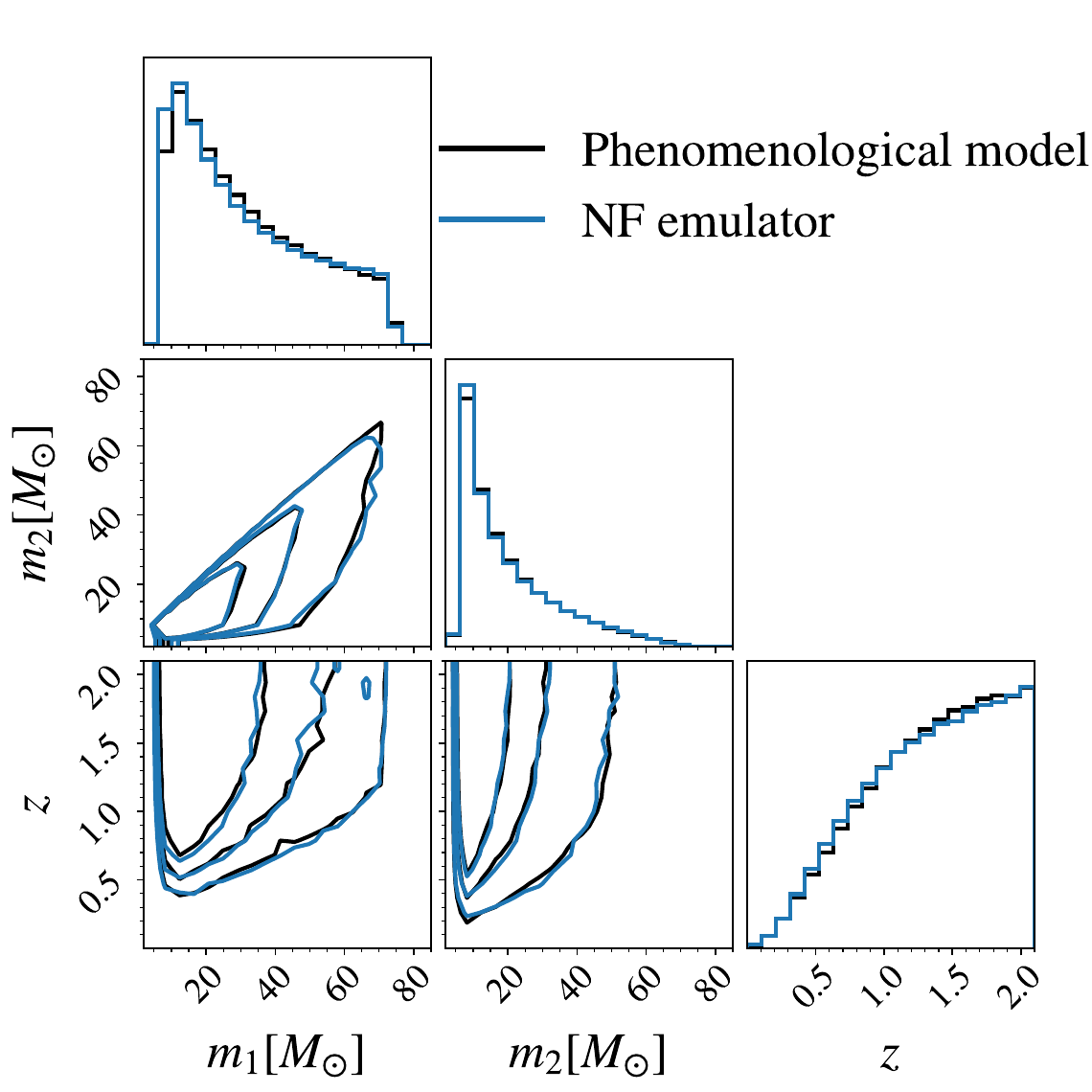}
    \includegraphics[width=0.45\textwidth]{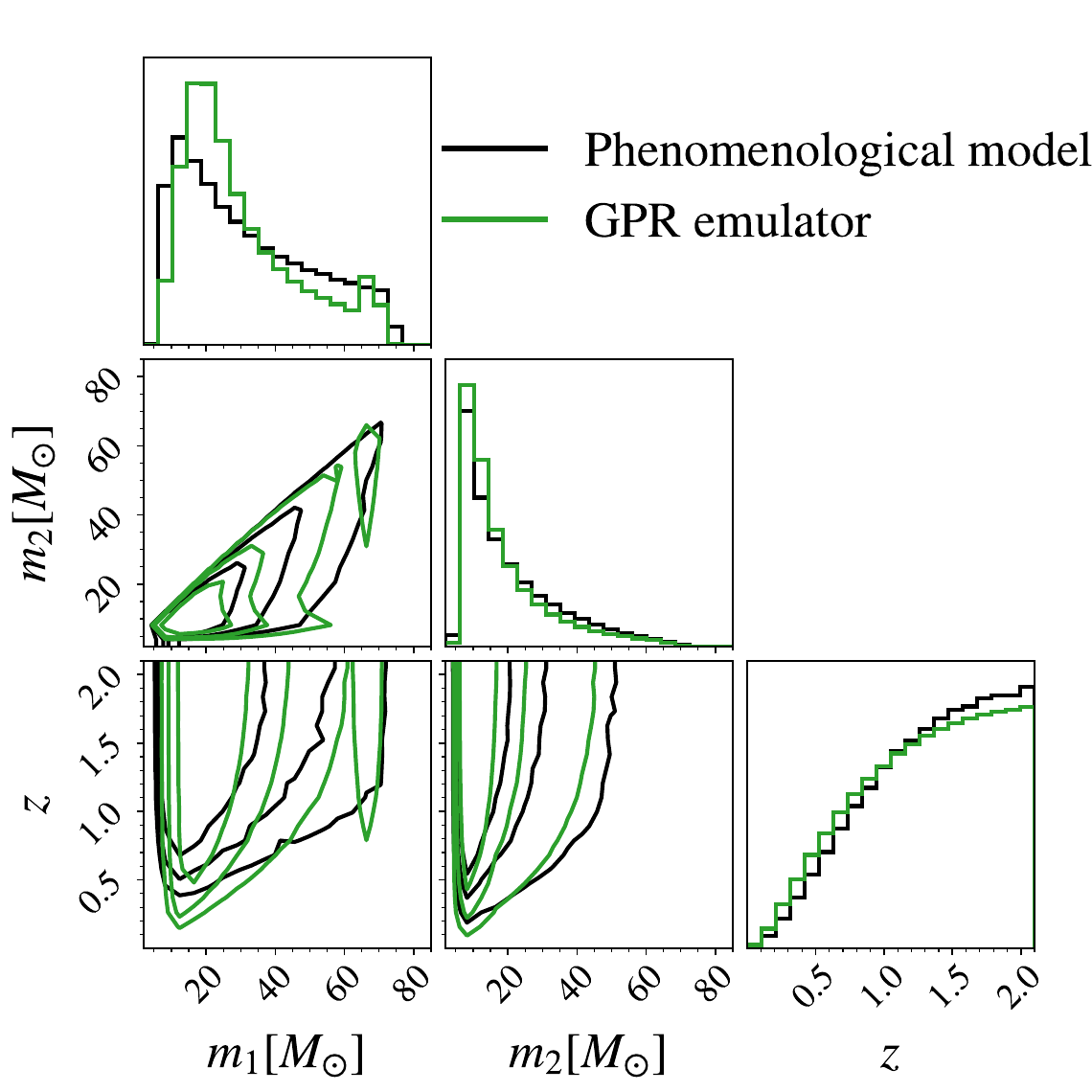}
    \caption{Test of the NF emulator and the GPR emulator. The test hyper-parameter $[ \alpha,\beta,m_{\mathrm{min}},m_{\mathrm{max}} ]=[3.0,0.5,6.0,74.0]$ is not included in the training set and validation set. The black curves show the events sampled by using the phenomenological model (true distributions), the blue curves (left panel) show those by using the NF emulator, and the green curves (right panel) show those by using the GPR emulator. The diagonal plots are marginalized distributions for each event parameter while the off-diagonal plots are the joint distributions between the event parameters. The three contour levels represent the 50\%, 70\%, and 90\% credible regions of the distributions. }
    \label{fig:dis} 
\end{figure*}
\section{Result}\label{result}
\subsection{Comparison on event distribution}

We compare the performance of emulators on sampling GW events to the phenomenological model.
In Fig. \ref{fig:dis}, we show the event distributions predicted by two emulators with test hyper-parameter $[ \alpha,\beta,m_{\mathrm{min}},m_{\mathrm{max}} ]=[ 3.0,0.5,6.0,74.0]$.
The event distributions predicted by the phenomenological model represent the true event distributions. 
The distributions predicted by the GPR emulator have significant discrepancies with the true event distributions.
In marginalized $m_1$ distribution, the one predicted by the GPR emulator does not agree with the true event distributions where the largest discrepancies appear at the mass limits.
Furthermore, the joint distributions of $m_1$ have an irregular shape when compared to the true event distributions.
Although the marginalized distribution of $m_2$ and $z$ have low discrepancies, the joint distribution is still significantly different from the true event distributions.
On the other hand, both the marginalized and joint distributions predicted by the NF emulator match the true distributions.
It can also recover the truncation characteristic at the limit of the event parameters.
The distribution similarity between using the NF/GPR emulator and the phenomenological model, as quantified by the Kullback-Leibler divergence \cite{KLdiv}, is $D_{\mathrm{KL}}=0.141, 0.277$ nat respectively.
A smaller $D_{\mathrm{KL}}$ indicates that two distributions are more similar, implying that the NF emulator's event distributions are more similar to the true distributions than the GPR emulator.

\subsection{Inference on mock data}
\begin{figure}
    \includegraphics[width=0.5\textwidth]{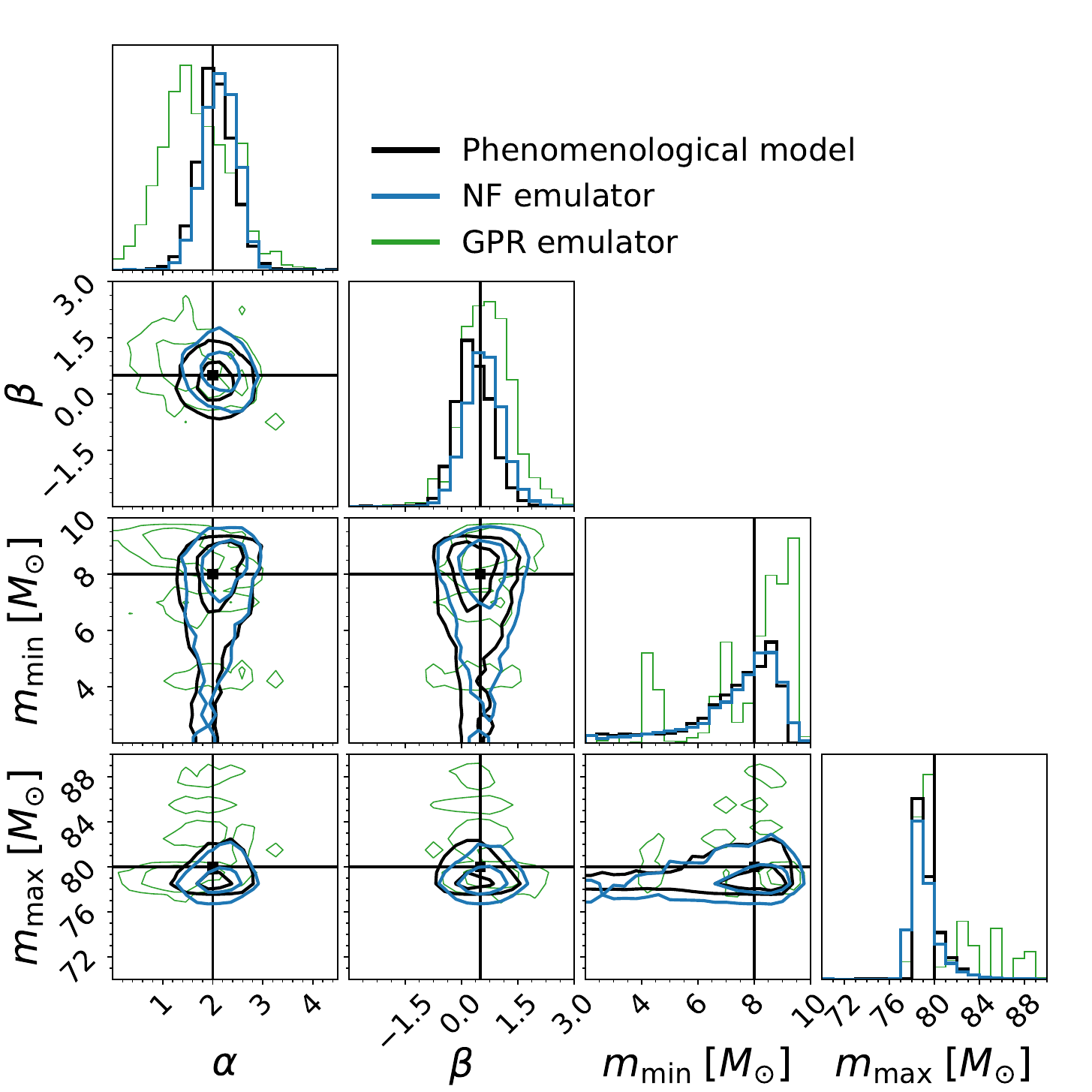}
    \caption{The sampled posterior distribution after injecting 50 events from a mock catalog with selection bias. The hyperparameters that characterize the mock catalog $[ \alpha,\beta,m_{\mathrm{min}},m_{\mathrm{max}} ]= [ 2.0,0.5,8.0,80.0 ]$ (true answers) are marked by the black lines. The black curves show the sampled posterior distributions by using the phenomenological model (true posterior distributions), the blue curves show those by using the NF emulator, and the green curves represent those by using the GPR emulator. The two contour levels represent the 50\% and 90\% credible regions of the distributions. The posterior distribution obtained by using the GPR emulator can barely recover the true posterior distribution. Even after training the GPR emulator with more data, the distributions remain to scatter and diverge. In contrast, the NF emulator recover the true posterior distribution.}
    \label{fig:mock_inj}
\end{figure}
\begin{figure*}
    \includegraphics[width=0.65\textwidth]{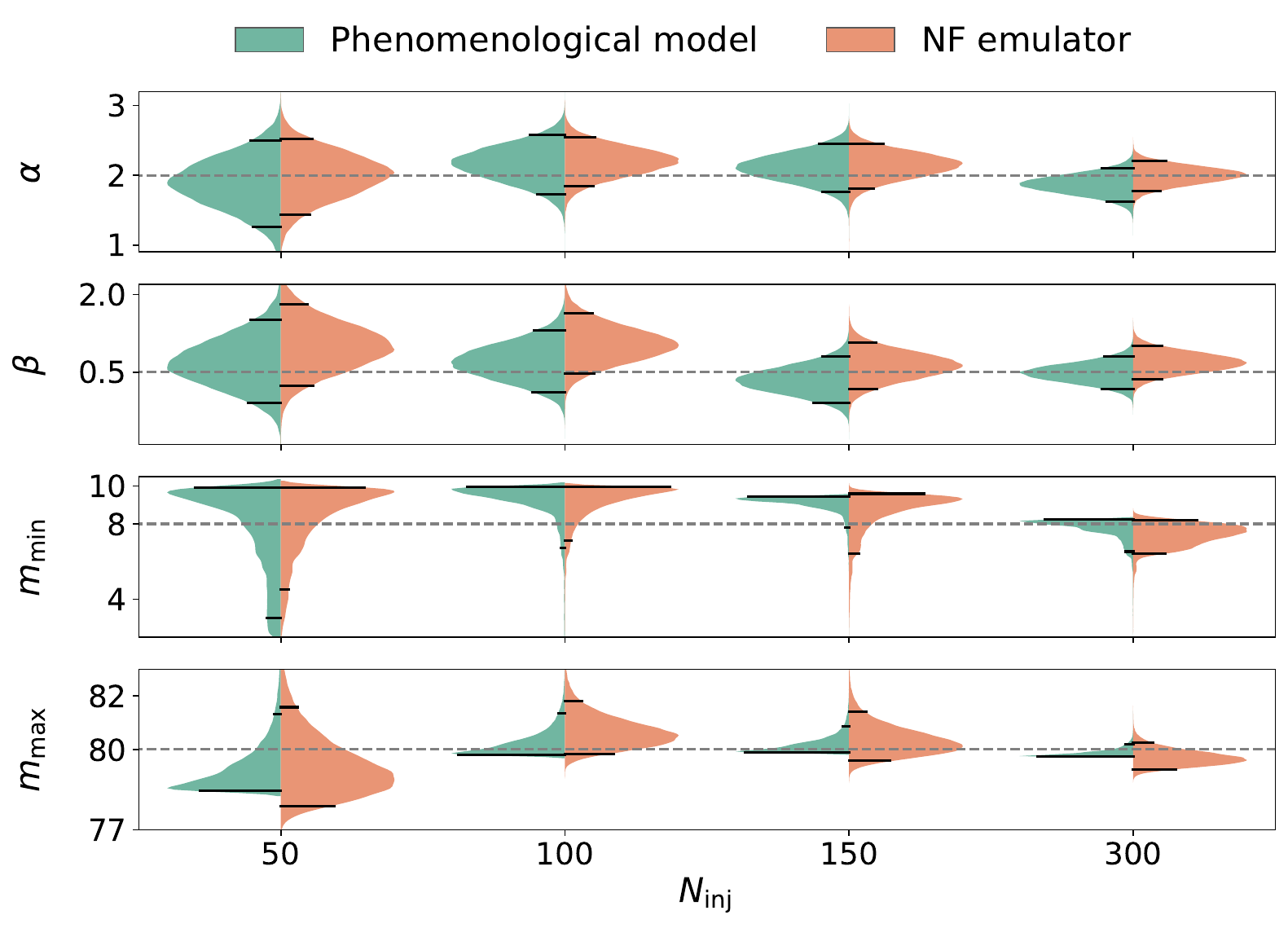}
    \caption{Marginalized posterior distribution on inferred hyperparameters by using the phenomenological model (green) and the NF emulator (orange). Inferences on different numbers of injections $N_{\mathrm{inj}}=50,100,150,300$ are performed. Horizontal black ticks and dashed grey lines mark the 90\% credible regions and true answers. The inferred hyperparameters by using the NF emulator agree with the true answers with 90\% credible region up to 300 injections.}
    \label{fig:violin plot}
\end{figure*}
Next, we examine the performance of the emulators on a population level.
To begin with, we build a mock catalog with the truncated power law model by using rejection sampling.
The hyperparameters that characterize it (true hyperparameters) are $[ \alpha,\beta,m_{\mathrm{min}},m_{\mathrm{max}} ] = [2.0,0.5,8.0,80.0]$.
Then, we evaluate the performance of two emulators by injecting 50 GW events from the mock catalog with selection bias. 
The sampled posterior distributions by using the phenomenological model represent the true posterior distributions. 

In Fig. \ref{fig:mock_inj}, we show the joint and marginalized posterior distributions of the hyper-parameter that favor the injected 50 GW events.
The sampled posterior distributions by using the GPR emulator diverge and scatter with only 50 GW events.
They have multiple local minimums which are different from the true posterior distributions.
The result suggests that the GPR emulator is not capable to act as a likelihood emulator even at a low-injection regime.
On the other hand, using the NF emulator can recover the true posterior distribution with low discrepancy.
They agree with each other in both marginalized distributions, joint distributions, and the most probable values.

As the number of observed GW events is rapidly increasing, the bias of inferred hyperparameters in simulation-based inference will become significant.
Therefore, we evaluate the performance of the NF emulator on more injections and compare with the phenomenological model.
We draw $50,100,150,300$ GW events from the mock catalog and infer the hyperparameters.
Figure \ref{fig:violin plot} shows the violin plots of marginalized posterior distribution of inferred hyperparameters by using the phenomenological model and the NF emulator.
Although the NF emulator's posterior distribution does not perfectly match that of the phenomenological model, the inferred hyperparameters agree with the true answers with 90\% credible regions up to 300 injections.
The marginalized posterior distribution of inferred hyperparameters by using the phenomenological model converges toward the true answer as the number of injections increases. 
On the other hand, the NF emulator gives similar convergence behavior.
The result shows the NF emulator is still a capable likelihood estimator when $N_{\mathrm{inj}}=300$ and selection bias is included.
However, notice that we use a smooth model (NF) to interpolate a hard cutoff phenomenological model. 
Therefore, the poor $m_{\mathrm{min}}/m_{\mathrm{max}}$ convergence performance is inevitable.
\begin{figure}
    \centering
    \includegraphics[width=0.4\textwidth]{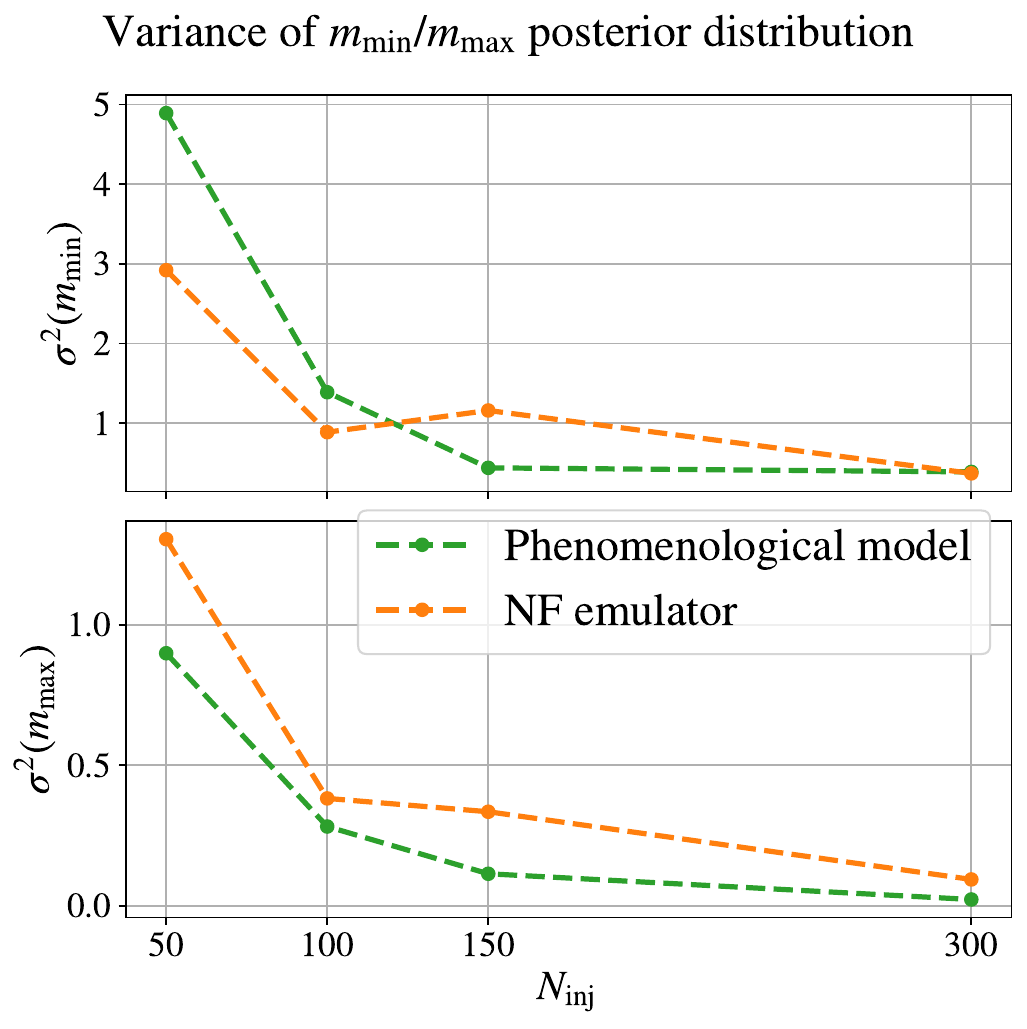}
    \caption{Variance of inferred $m_{\mathrm{min}},m_{\mathrm{max}}$ posterior distribution in Fig. \ref{fig:violin plot}. The upper/bottom panel shows the variance of inferred $m_{\mathrm{min}}/m_{\mathrm{max}}$ posterior distribution. Green lines represent the sampled posterior distributions by using the phenomenological model while orange lines represent those by using the NF emulator. The $m_{\mathrm{max}}$ uncertainty shrinks slower when using the NF emulator.}
    \label{fig:violin variance}
\end{figure}
Figure \ref{fig:violin variance}, shows the variance of inferred $m_{\mathrm{min}}/m_{\mathrm{max}}$ posterior distribution against the number of injections. 
When compared to the phenomenological model, the $m_{\mathrm{max}}$ uncertainty shrinks slower when using the NF emulator, but there is no big difference for the $m_{\mathrm{min}}$ uncertainty. 
Because we have more observed events near $m_{\mathrm{max}}$ than $m_{\mathrm{min}}$, the limitation of the NF emulator is more clearly shown in the inferred $m_{\mathrm{max}}$ posterior distribution.

\subsection{Data from GWTC-2}
\begin{figure*}
    \centering
    \includegraphics[width=0.4\textwidth]{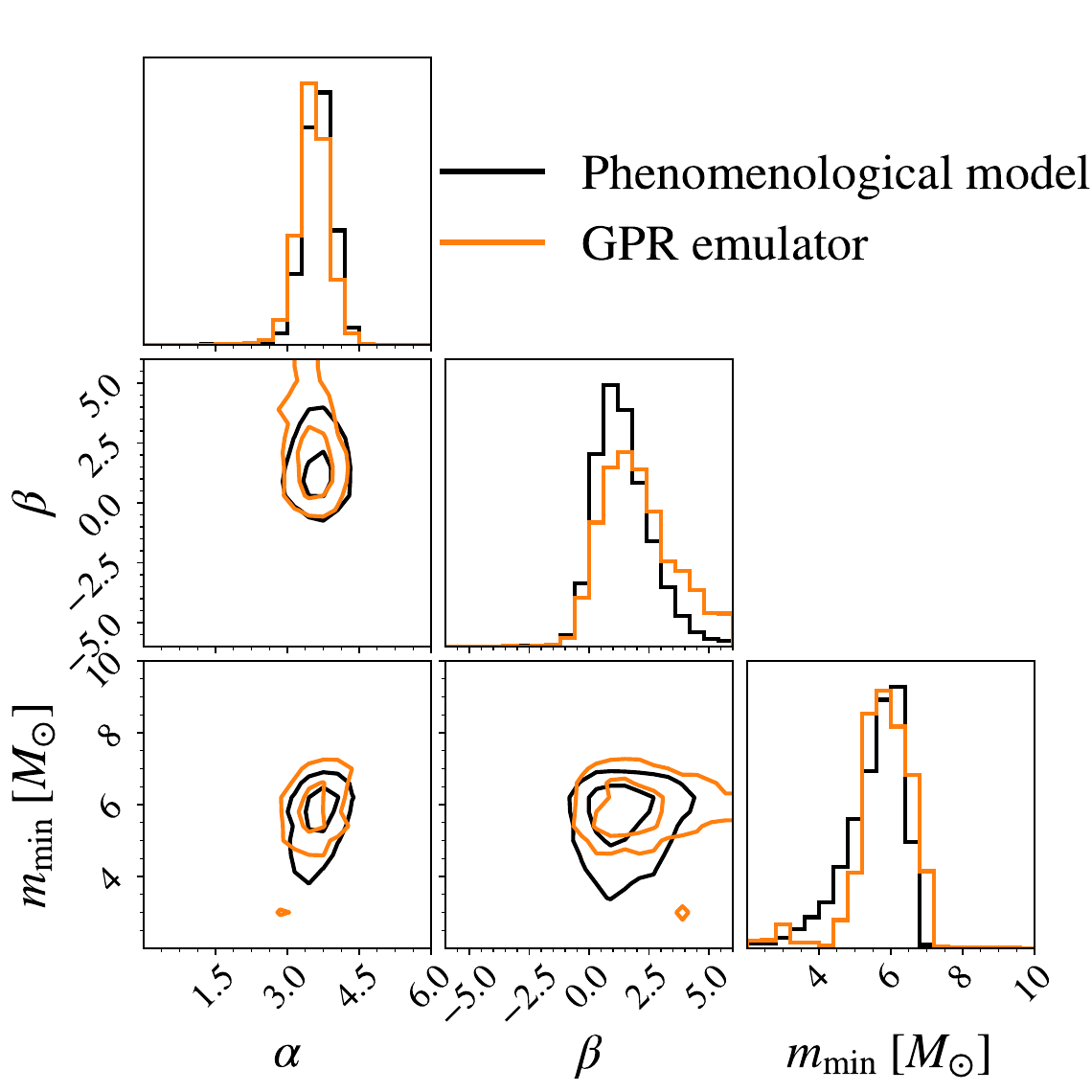}
    \includegraphics[width=0.5\textwidth]{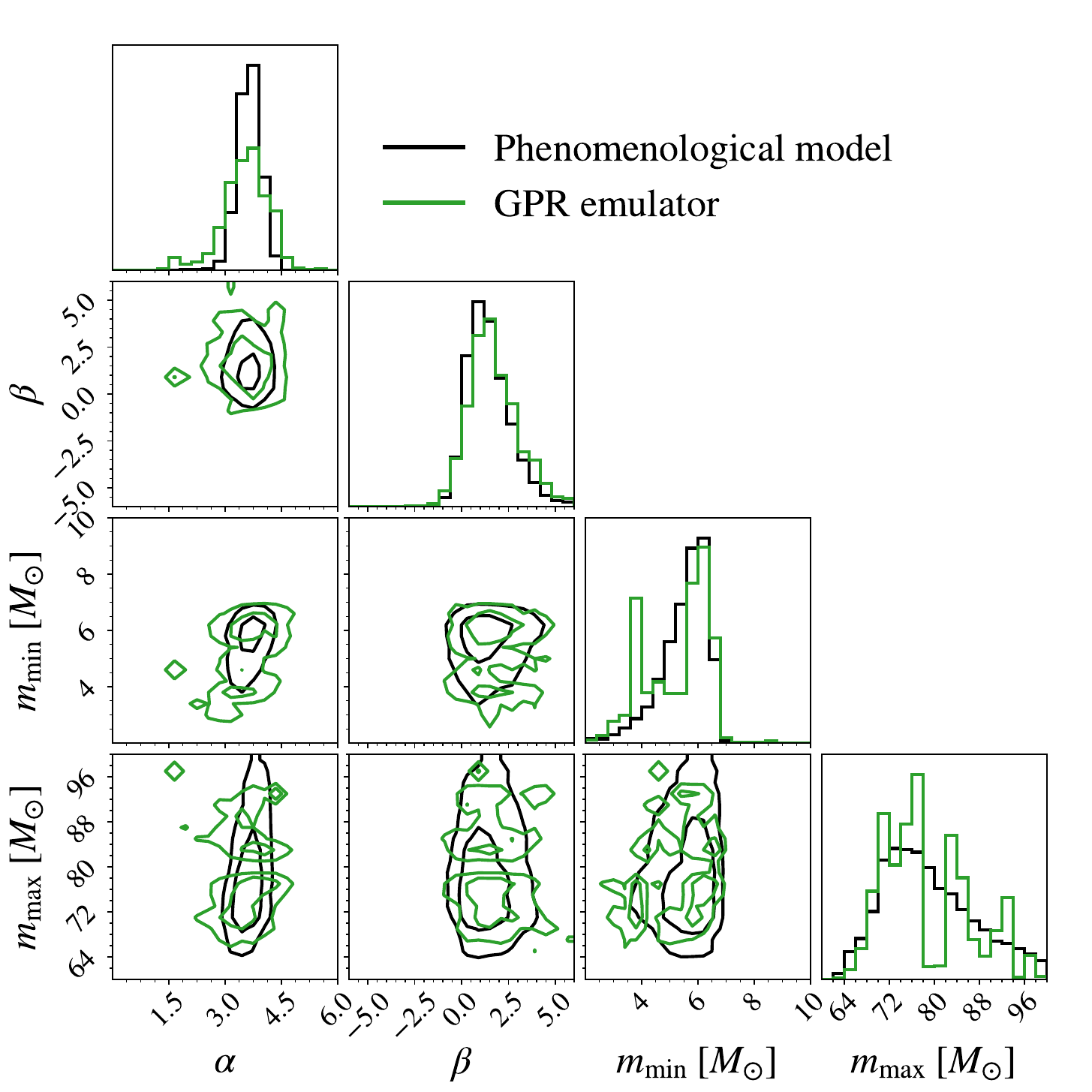}
    \caption{Comparison of sampled posterior distributions on GWTC-2 data by using the phenomenological model and the GPR emulator. The black curves represent the sampled posterior distributions by using the phenomenological model, the orange curves (left) represent those by using the GPR emulator trained on only three hyperparameters:$[\alpha,\beta ,m_{\mathrm{min}}]$, while the green curves (right) are those by using the GPR emulator trained on all four hyperparameters:$[\alpha,\beta ,m_{\mathrm{min}},m_{\mathrm{max}}]$. The two contour levels represent the 50\% and 90\% credible regions of the sampled posterior distribution. For three hyperparameters, the GPR emulator can reconstruct the distribution using the phenomenological model, but not for four hyperparameters. Even after training the GPR emulator with more data, the distributions persist to scatter and diverge.  }
    \label{fig:gpr_gwtc}
\end{figure*}
\begin{figure}
    \centering
    \includegraphics[width=0.5\textwidth]{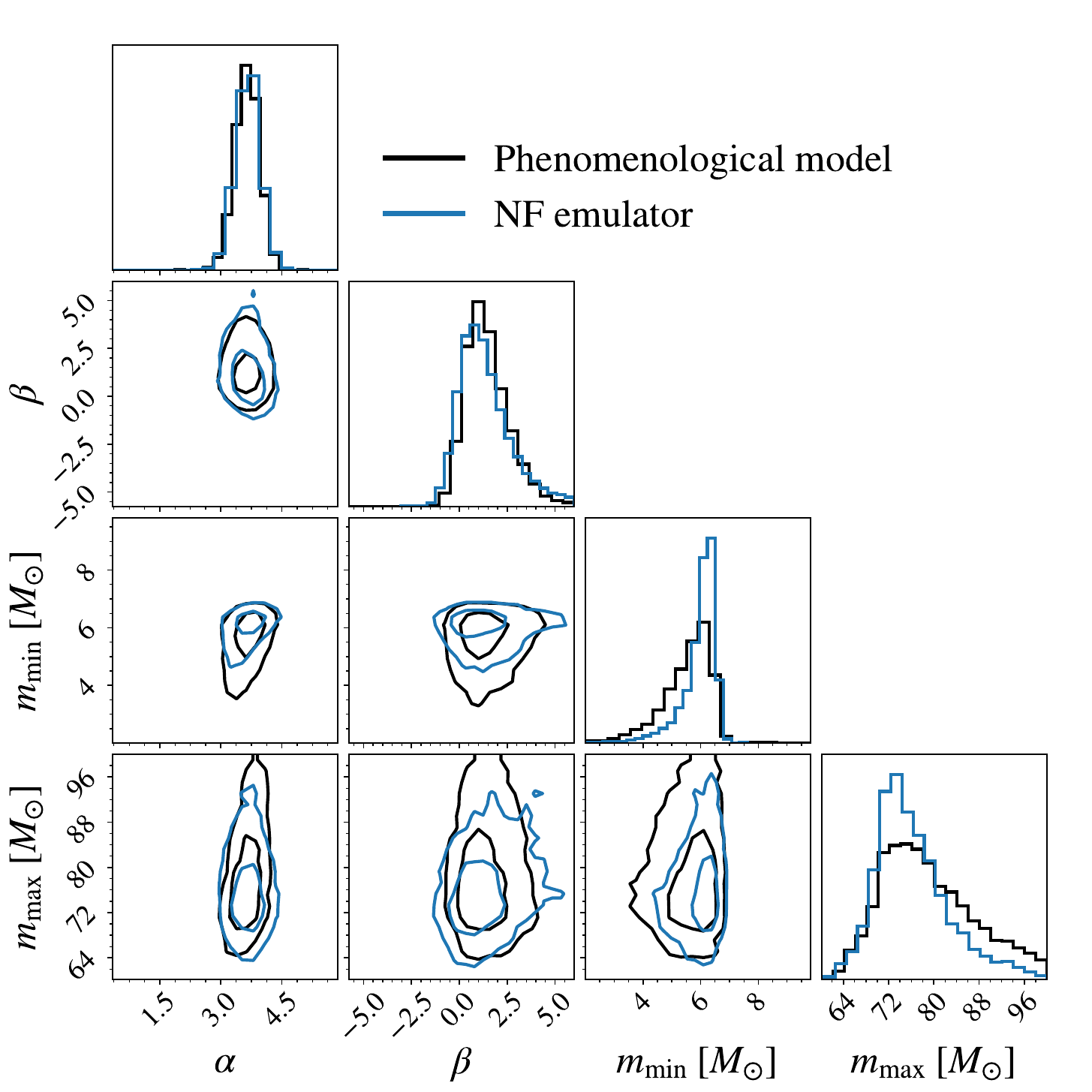}
    \caption{Comparison of sampled posterior distributions on GWTC-2 data by using the phenomenological model and NF emulator. The black curves indicate the sampled posterior distributions by using the phenomenological model while the blue curves represent those by using the NF emulator. The two contour levels represent the 50\% and 90\% credible regions. The NF emulator can recover the distribution by using the phenomenological model except for the occurrence of underestimated uncertainty in the $m_{\mathrm{min}}$ and $m_{\mathrm{max}}$ distributions.}
    \label{fig:nf_gwtc}
\end{figure}
We also evaluate the performance of the emulator on real data. 
For simplicity, we only use $44$ high-significance events from GWTC-2 \cite{GWTC2} because our goal is to compare the emulators' performance to the phenomenological model.
The dataset is the same subset chosen for the population analyses in \cite{POPgwtc2}.
In particular, we exclude three events with a large false-alarm rate (GW190426, GW190719, GW190909) and three events with $m_2<3M_{\odot}$ (GW170817,
GW190425, GW190814).

We performed two population inferences on GWTC-2 data by using the GPR emulator as shown in Fig. \ref{fig:gpr_gwtc}.
One only trained the GPR emulator on three hyperparameter $[\alpha,\beta ,m_{\mathrm{min}}]$ (left panel), another one trained on all hyperparameters (right panel).
Since those hyperparameters are independent of each other, the result of inferring three or four hyperparameters will be the same. 
For inferring three hyperparameters, the GPR emulator can recover the posterior distribution by using the phenomenological model with fair convergence. 
In the four hyperparameters cases, they have irregular and diverge distributions, which do not agree with the sampled distributions obtained by using the phenomenological model.

The performance of the population inference on GWTC-2 data by using the NF emulator is shown in Fig. \ref{fig:nf_gwtc}.
The NF emulator is capable of recovering the sampled distribution by using the phenomenological model except for the $m_{\mathrm{min}}$ and $m_{\mathrm{max}}$ related distributions.
Although their most probable values are aligned, the uncertainties predicted by the NF emulator are smaller for $m_{\mathrm{min}}$ and $m_{\mathrm{max}}$. 
The result shows the NF emulator can not learn the truncation characteristic perfectly.

\section{Discussion} \label{sec:discussion}

\begin{figure}[H]
    \includegraphics[width=0.4\textwidth]{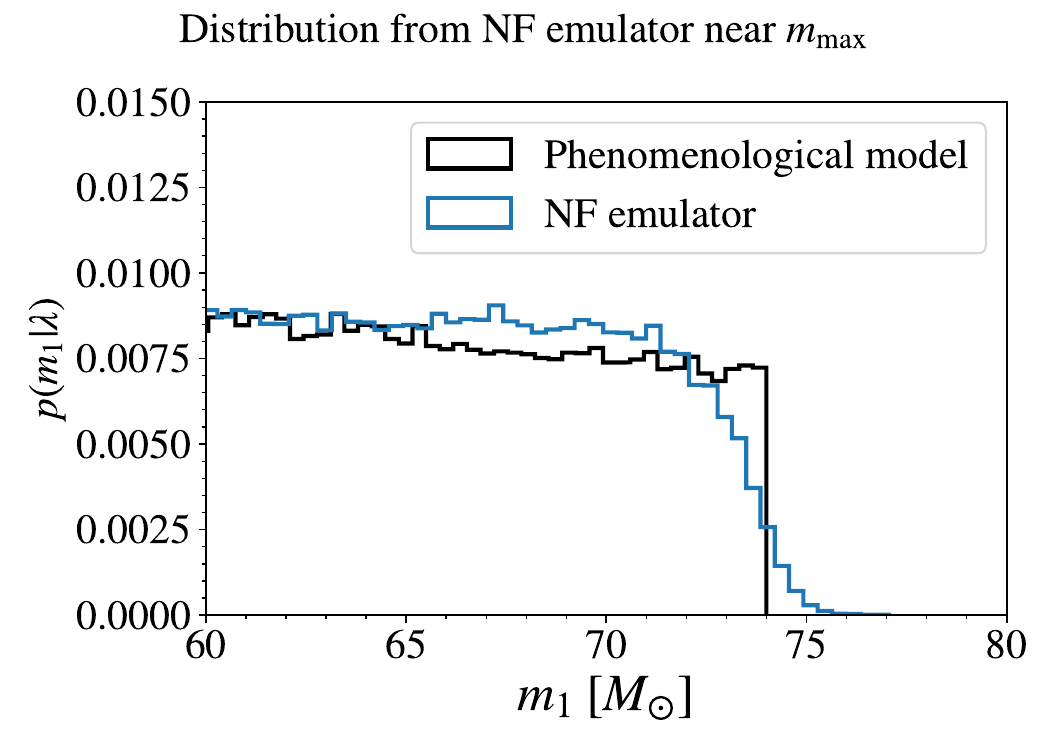}
    \caption{The marginalized $m_1$ distributions near $m_{\mathrm{max}}$ on event sampling. The test hyper-parameter are $[ \alpha,\beta,m_{\mathrm{min}},m_{\mathrm{max}} ]=[3.0,0.5,6.0,74.0]$ which approximately equal to the most probable inferred hyperparameters in Fig. \ref{fig:nf_gwtc}. The black curves represent the marginalized probability density predicted by the phenomenological model while blue curves present those by the NF emulator. The NF emulator can not capture the truncation perfectly. }
    \label{fig:nf_discrepancy}
\end{figure}

We showed the performance of the GPR and NF emulators by employing a truncated power law model. 
For the GPR emulator, it is hard to sample the correct event distribution near the truncation as shown in Fig. \ref{fig:dis}.
In addition, the sampled posterior distribution of $m_{\mathrm{min}},m_{\mathrm{max}}$ scatter more strongly than $\alpha$ and $\beta$ in Fig. \ref{fig:mock_inj}. 
The results reveal the inability of the GPR emulator to learn the truncation property. 
For a truncated power law distribution, we have relatively fewer data near the truncation.
However, we need relatively more data to learn the sharp edge. 
As a result, we have insufficient training data to learn the truncation property and thus have poor performance on the population inference.
Moreover, we need to specify the number of bins and bin width to train a GPR emulator.
It introduces the Poisson uncertainty in each bin which is $\propto \frac{1}{\sqrt{N}}$, where $N$ is the number of events in that bin.
The number of events in some bins may equal zero even if the theoretical probability density is not zero; consequently, it will produce a large Poisson uncertainty.
A sufficiently large number of events is needed to recover the theoretical probability density before training. 
For high-dimensional cases, the number of events needed to recover the theoretical joint probability density grows exponentially.
To solve the problem, one possible approach is to divide the event parameter space into two regions—region 1 with plenty of samples and region 2 with fewer samples.
Then we can continue drawing samples in region 2 until we have enough. 
After that, we can construct the GPR emulator with the reweighted samples.
However, even if we used theoretical probability density, the number of bins still affects the resolution of the density estimation.
The comparison shown in Fig. \ref{fig:gpr_gwtc} demonstrates the inability of the GPR emulator on higher dimensions.
We tried using more simulations as training data with the higher binning resolution, but the scatter and diverge problem persisted.
The largest training dataset we tried took around a week to train using a 4-core CPU.
Not to mention the time spent on generating the training data.
As a result, training a good GPR emulator for population inference is unaffordable in terms of time and computational cost.
The result may reveal GPR's limitations in estimating high-dimensional density \cite{liugprlimitation,2020gprlimitation}.
GPR learns the model by inferring training data with a Gaussian prior. 
As a result, the predicted likelihood function will look like a sine curve that connects the training data; it has many local minimums and provides an explanation for the scattered posterior distribution in Fig. \ref{fig:mock_inj} and Fig. \ref{fig:gpr_gwtc}.
On the other hand, the NF emulator performs well except for underestimating the uncertainty for some hyper-parameter as shown in Fig. \ref{fig:nf_gwtc}. But the uncertainty estimated by the emulator should be greater than the phenomenological model because of the limited training data. 
The problem of the underestimated uncertainty may come from the nature of the NF emulator \cite{SBI_crisis}.
NF is a series of continuous transformations so has relatively bad performance on learning truncation property.
At the truncation of distribution, NF prefers a smooth change rather than a sharp truncation as shown in Fig. \ref{fig:nf_discrepancy} at $m_1 =m_{\mathrm{max}}$.
It requires infinitely many continuous transformations to get a sharp truncation. 
In addition, the training depends on the loss function which is the likelihood of the entire transformation.
The loss function takes care of the entire training data at the same time so that it is unlikely to have binning and resolution problems.
As a result, the uncertainty near the truncation will smooth out. 
Studies on controlling the uncertainty accumulated in the training should be carried out.

After understanding the characteristics of the emulator, we should use with caution when applying the techniques to those state-of-art models \cite{m1,m2,m3,m4} for more sophisticated GW population studies.  
We can use the technique to eliminate the synthesis simulations which are not favored by the observation data by marginalizing the population posteriors from two models and computing the Bayes factor between the two models \cite{para_estimation,model_selection_bayes}.
However, some simulations may be ruled out wrongly because of the underestimated uncertainty behavior of the NF emulator.  

Furthermore, with the fast growth of the GW observed catalog, one potential research direction is to train an emulator using real GW data to sample the real GW distribution without employing any population model.
However, each GW event is expressed as the posterior samples from the parameter estimation because of the measurement uncertainty.
Machine learning frequently fails to handle such type of training data. 
In particular, the technique we used in the NF emulator is not regulated to train by using such type of data.
One approach is to construct the Bayesian neural network \cite{NF_BNN} with NF.
Studies and performance tests of this technique in GW population analysis should be carried out in the future.

\section{ACKNOWLEDGMENTS}
K.W.K.W. and S.H. are supported by the Simons Foundation.
O.A.H. was partially supported by grants from the Research Grants Council of the Hong Kong (Project No. CUHK 14306218), The Croucher Foundation of Hong Kong and Research Committee of the Chinese University of Hong Kong. 
T.G.F.L was partially supported by grants from the Research Grants Council of the Hong Kong (Project No. CUHK 14306419), The Croucher Foundation of Hong Kong and Research Committee of the Chinese University of Hong Kong. 
This project was conducted using computational resources at the Rusty cluster of the Flatiron Institute. 

This project made use of the following Python packages: \textbf{Matplotlib} \cite{matplotlib}, \textbf{NumPy} \cite{numpy}, \textbf{SciPy} \cite{scipy}, \textbf{scikit-learn} \cite{scikitlearn}, \textbf{corner} \cite{corner}, \textbf{emcee} \cite{corner} and \textbf{PyTorch} \cite{torch}. 

This material is based upon work supported by NSF's LIGO Laboratory which is a major facility fully funded by the National Science Foundation.
This research has made use of data, software and/or
web tools obtained from the GW Open Science Center (https://www.gw-openscience.org), a service of LIGO Laboratory, the LIGO Scientific Collaboration and the Virgo Collaboration.
Virgo is funded by the French Centre National de Recherche Scientifique (CNRS), the Italian Istituto Nazionale della Fisica Nucleare (INFN) and the Dutch Nikhef, with contributions by Polish and
Hungarian institutes.

%\onecolumngrid
\appendix
\section{Mock Data Catalog}

\label{A1}
Here is the detail about the formulas used to generate mock Gravitational-wave events catalog.\par 
First, the distribution of $m_1$ follows a power law distribution with spectral index $\alpha$:
\begin{equation}
\label{eq:9.1}
p(m_1 |\alpha,m_{\mathrm{min}},m_{\mathrm{max}}) \propto \left\{ \begin{array}{ll}

{m_1}^{-\alpha} & \quad m_{\mathrm{min}}<m_1 <m_{\mathrm{max}} \\
0 \quad & \text{otherwise}.
\end{array} \right.
\end{equation}

Second, the mass ration $q=m_2/m_1$ follows a power law distribution with spectral index $\beta$:

\begin{equation}
\label{eq:9.2}
p(q|\beta,m_{\mathrm{min}},m_1) \propto \left\{ \begin{array}{ll}
{q}^{\beta} & \quad m_{\mathrm{min}}< m_2 < m_1 \\
0 & \quad \text{otherwise}.
\end{array} \right.
\end{equation}

And lastly, the red-shift distribution can be written as:
\begin{equation}
\label{eq:9.3}
    p(z) \propto (1+z)^{\kappa -1} \frac{dV_c}{dz}\quad z \in [0,2.3],
\end{equation}

where $\kappa$ is the redshift evolution parameter and is set to 1, and $dV_c / dz$ is the differential comving volume.

\begin{table*}
\centering
\begin{tabular}{c c c c} % centered columns (4 columns)
\hline\hline %inserts double horizontal lines
Hyper-parameter & Description & Prior\\ [0.8ex] % inserts table
%heading
\hline % inserts single horizontal line

$\alpha$ & power law index on $m_1$ & U(-6,6) \\
$\beta$ & power law index on $q=m_1/m_2$ & U(-6,6) \\
$m_{\mathrm{min}}$ & Minimum mass for mass distribution $m_1$ & U(2$M_{\odot}$,10 $M_{\odot}$)\\
$m_{\mathrm{max}}$ & Maximum mass for mass distribution $m_1$ & U(60$M_{\odot}$,100$M_{\odot}$)\\ [1ex] % [1ex] adds vertical space
\hline %inserts single line
\end{tabular}
\caption{Summary of truncated power law hyperparameters.} % title of Table
\label{table:hpyer-parameter} % is used to refer this table in the text
\end{table*}
\bibliography{testrobustness}

%merlin.mbs apsrev4-1.bst 2010-07-25 4.21a (PWD, AO, DPC) hacked
%Control: key (0)
%Control: author (8) initials jnrlst
%Control: editor formatted (1) identically to author
%Control: production of article title (-1) disabled
%Control: page (0) single
%Control: year (1) truncated
%Control: production of eprint (0) enabled
\begin{thebibliography}{66}%
\makeatletter
\providecommand \@ifxundefined [1]{%
 \@ifx{#1\undefined}
}%
\providecommand \@ifnum [1]{%
 \ifnum #1\expandafter \@firstoftwo
 \else \expandafter \@secondoftwo
 \fi
}%
\providecommand \@ifx [1]{%
 \ifx #1\expandafter \@firstoftwo
 \else \expandafter \@secondoftwo
 \fi
}%
\providecommand \natexlab [1]{#1}%
\providecommand \enquote  [1]{``#1''}%
\providecommand \bibnamefont  [1]{#1}%
\providecommand \bibfnamefont [1]{#1}%
\providecommand \citenamefont [1]{#1}%
\providecommand \href@noop [0]{\@secondoftwo}%
\providecommand \href [0]{\begingroup \@sanitize@url \@href}%
\providecommand \@href[1]{\@@startlink{#1}\@@href}%
\providecommand \@@href[1]{\endgroup#1\@@endlink}%
\providecommand \@sanitize@url [0]{\catcode `\\12\catcode `\$12\catcode
  `\&12\catcode `\#12\catcode `\^12\catcode `\_12\catcode `\%12\relax}%
\providecommand \@@startlink[1]{}%
\providecommand \@@endlink[0]{}%
\providecommand \url  [0]{\begingroup\@sanitize@url \@url }%
\providecommand \@url [1]{\endgroup\@href {#1}{\urlprefix }}%
\providecommand \urlprefix  [0]{URL }%
\providecommand \Eprint [0]{\href }%
\providecommand \doibase [0]{http://dx.doi.org/}%
\providecommand \selectlanguage [0]{\@gobble}%
\providecommand \bibinfo  [0]{\@secondoftwo}%
\providecommand \bibfield  [0]{\@secondoftwo}%
\providecommand \translation [1]{[#1]}%
\providecommand \BibitemOpen [0]{}%
\providecommand \bibitemStop [0]{}%
\providecommand \bibitemNoStop [0]{.\EOS\space}%
\providecommand \EOS [0]{\spacefactor3000\relax}%
\providecommand \BibitemShut  [1]{\csname bibitem#1\endcsname}%
\let\auto@bib@innerbib\@empty
%</preamble>
\bibitem [{\citenamefont {Abbott}\ \emph {et~al.}(2016)\citenamefont {Abbott},
  \citenamefont {Abbott}, \citenamefont {Abbott}, \citenamefont {Abernathy},
  \citenamefont {Acernese}, \citenamefont {Ackley}, \citenamefont {Adams},
  \citenamefont {Adams}, \citenamefont {Addesso}, \citenamefont {Adhikari},\
  and\ \citenamefont {et~al.}}]{firstGW}%
  \BibitemOpen
  \bibfield  {author} {\bibinfo {author} {\bibfnamefont {B.}~\bibnamefont
  {Abbott}}, \bibinfo {author} {\bibfnamefont {R.}~\bibnamefont {Abbott}},
  \bibinfo {author} {\bibfnamefont {T.}~\bibnamefont {Abbott}}, \bibinfo
  {author} {\bibfnamefont {M.}~\bibnamefont {Abernathy}}, \bibinfo {author}
  {\bibfnamefont {F.}~\bibnamefont {Acernese}}, \bibinfo {author}
  {\bibfnamefont {K.}~\bibnamefont {Ackley}}, \bibinfo {author} {\bibfnamefont
  {C.}~\bibnamefont {Adams}}, \bibinfo {author} {\bibfnamefont
  {T.}~\bibnamefont {Adams}}, \bibinfo {author} {\bibfnamefont
  {P.}~\bibnamefont {Addesso}}, \bibinfo {author} {\bibfnamefont
  {R.}~\bibnamefont {Adhikari}}, \ and\ \bibinfo {author} {\bibnamefont
  {et~al.}},\ }\href {\doibase 10.1103/physrevd.93.122003} {\bibfield
  {journal} {\bibinfo  {journal} {Physical Review D}\ }\textbf {\bibinfo
  {volume} {93}} (\bibinfo {year} {2016}),\
  10.1103/physrevd.93.122003}\BibitemShut {NoStop}%
\bibitem [{\citenamefont {Aasi}\ \emph {et~al.}(2015)\citenamefont {Aasi},
  \citenamefont {Abbott}, \citenamefont {Abbott}, \citenamefont {Abbott},
  \citenamefont {Abernathy}, \citenamefont {Ackley}, \citenamefont {Adams},
  \citenamefont {Adams}, \citenamefont {Addesso}, \citenamefont {Adhikari},
  \citenamefont {Adya}, \citenamefont {Affeldt}, \citenamefont {Aggarwal},
  \citenamefont {Aguiar}, \citenamefont {Ain}, \citenamefont {Ajith},
  \citenamefont {Alemic}, \citenamefont {Allen}, \citenamefont {Amariutei},
  \citenamefont {Anderson}, \citenamefont {Anderson}, \citenamefont {Arai},
  \citenamefont {Araya}, \citenamefont {Arceneaux}, \citenamefont {Areeda},
  \citenamefont {Ashton}, \citenamefont {Ast}, \citenamefont {Aston},
  \citenamefont {Aufmuth}, \citenamefont {Aulbert}, \citenamefont {Aylott},
  \citenamefont {Babak}, \citenamefont {Baker}, \citenamefont {Ballmer},
  \citenamefont {Barayoga}, \citenamefont {Barbet}, \citenamefont {Barclay},\
  and\ \citenamefont {et~al.}}]{AdvancedLIGO}%
  \BibitemOpen
  \bibfield  {author} {\bibinfo {author} {\bibfnamefont {J.}~\bibnamefont
  {Aasi}}, \bibinfo {author} {\bibfnamefont {B.~P.}\ \bibnamefont {Abbott}},
  \bibinfo {author} {\bibfnamefont {R.}~\bibnamefont {Abbott}}, \bibinfo
  {author} {\bibfnamefont {T.}~\bibnamefont {Abbott}}, \bibinfo {author}
  {\bibfnamefont {M.~R.}\ \bibnamefont {Abernathy}}, \bibinfo {author}
  {\bibfnamefont {K.}~\bibnamefont {Ackley}}, \bibinfo {author} {\bibfnamefont
  {C.}~\bibnamefont {Adams}}, \bibinfo {author} {\bibfnamefont
  {T.}~\bibnamefont {Adams}}, \bibinfo {author} {\bibfnamefont
  {P.}~\bibnamefont {Addesso}}, \bibinfo {author} {\bibfnamefont {R.~X.}\
  \bibnamefont {Adhikari}}, \bibinfo {author} {\bibfnamefont {V.}~\bibnamefont
  {Adya}}, \bibinfo {author} {\bibfnamefont {C.}~\bibnamefont {Affeldt}},
  \bibinfo {author} {\bibfnamefont {N.}~\bibnamefont {Aggarwal}}, \bibinfo
  {author} {\bibfnamefont {O.~D.}\ \bibnamefont {Aguiar}}, \bibinfo {author}
  {\bibfnamefont {A.}~\bibnamefont {Ain}}, \bibinfo {author} {\bibfnamefont
  {P.}~\bibnamefont {Ajith}}, \bibinfo {author} {\bibfnamefont
  {A.}~\bibnamefont {Alemic}}, \bibinfo {author} {\bibfnamefont
  {B.}~\bibnamefont {Allen}}, \bibinfo {author} {\bibfnamefont
  {D.}~\bibnamefont {Amariutei}}, \bibinfo {author} {\bibfnamefont {S.~B.}\
  \bibnamefont {Anderson}}, \bibinfo {author} {\bibfnamefont {W.~G.}\
  \bibnamefont {Anderson}}, \bibinfo {author} {\bibfnamefont {K.}~\bibnamefont
  {Arai}}, \bibinfo {author} {\bibfnamefont {M.~C.}\ \bibnamefont {Araya}},
  \bibinfo {author} {\bibfnamefont {C.}~\bibnamefont {Arceneaux}}, \bibinfo
  {author} {\bibfnamefont {J.~S.}\ \bibnamefont {Areeda}}, \bibinfo {author}
  {\bibfnamefont {G.}~\bibnamefont {Ashton}}, \bibinfo {author} {\bibfnamefont
  {S.}~\bibnamefont {Ast}}, \bibinfo {author} {\bibfnamefont {S.~M.}\
  \bibnamefont {Aston}}, \bibinfo {author} {\bibfnamefont {P.}~\bibnamefont
  {Aufmuth}}, \bibinfo {author} {\bibfnamefont {C.}~\bibnamefont {Aulbert}},
  \bibinfo {author} {\bibfnamefont {B.~E.}\ \bibnamefont {Aylott}}, \bibinfo
  {author} {\bibfnamefont {S.}~\bibnamefont {Babak}}, \bibinfo {author}
  {\bibfnamefont {P.~T.}\ \bibnamefont {Baker}}, \bibinfo {author}
  {\bibfnamefont {S.~W.}\ \bibnamefont {Ballmer}}, \bibinfo {author}
  {\bibfnamefont {J.~C.}\ \bibnamefont {Barayoga}}, \bibinfo {author}
  {\bibfnamefont {M.}~\bibnamefont {Barbet}}, \bibinfo {author} {\bibfnamefont
  {S.}~\bibnamefont {Barclay}}, \ and\ \bibinfo {author} {\bibnamefont
  {et~al.}},\ }\href {\doibase 10.1088/0264-9381/32/7/074001} {\bibfield
  {journal} {\bibinfo  {journal} {Classical and Quantum Gravity}\ }\textbf
  {\bibinfo {volume} {32}},\ \bibinfo {pages} {074001} (\bibinfo {year}
  {2015})}\BibitemShut {NoStop}%
\bibitem [{\citenamefont {Acernese}\ \emph {et~al.}(2014)\citenamefont
  {Acernese}, \citenamefont {Agathos}, \citenamefont {Agatsuma}, \citenamefont
  {Aisa}, \citenamefont {Allemandou}, \citenamefont {Allocca}, \citenamefont
  {Amarni}, \citenamefont {Astone}, \citenamefont {Balestri}, \citenamefont
  {Ballardin}, \citenamefont {Barone}, \citenamefont {Baronick}, \citenamefont
  {Barsuglia}, \citenamefont {Basti}, \citenamefont {Basti}, \citenamefont
  {Bauer}, \citenamefont {Bavigadda}, \citenamefont {Bejger}, \citenamefont
  {Beker}, \citenamefont {Belczynski}, \citenamefont {Bersanetti},
  \citenamefont {Bertolini}, \citenamefont {Bitossi}, \citenamefont {Bizouard},
  \citenamefont {Bloemen}, \citenamefont {Blom}, \citenamefont {Boer},
  \citenamefont {Bogaert}, \citenamefont {Bondi}, \citenamefont {Bondu},
  \citenamefont {Bonelli}, \citenamefont {Bonnand}, \citenamefont {Boschi},
  \citenamefont {Bosi}, \citenamefont {Bouedo}, \citenamefont {Bradaschia},
  \citenamefont {Branchesi}, \citenamefont {Briant}, \citenamefont {Brillet},
  \citenamefont {Brisson}, \citenamefont {Bulik}, \citenamefont {Bulten},
  \citenamefont {Buskulic}, \citenamefont {Buy}, \citenamefont {Cagnoli},
  \citenamefont {Calloni}, \citenamefont {Campeggi}, \citenamefont {Canuel},
  \citenamefont {Carbognani}, \citenamefont {Cavalier}, \citenamefont
  {Cavalieri},\ and\ \citenamefont {et~al.}}]{AdvancedVIRGO}%
  \BibitemOpen
  \bibfield  {author} {\bibinfo {author} {\bibfnamefont {F.}~\bibnamefont
  {Acernese}}, \bibinfo {author} {\bibfnamefont {M.}~\bibnamefont {Agathos}},
  \bibinfo {author} {\bibfnamefont {K.}~\bibnamefont {Agatsuma}}, \bibinfo
  {author} {\bibfnamefont {D.}~\bibnamefont {Aisa}}, \bibinfo {author}
  {\bibfnamefont {N.}~\bibnamefont {Allemandou}}, \bibinfo {author}
  {\bibfnamefont {A.}~\bibnamefont {Allocca}}, \bibinfo {author} {\bibfnamefont
  {J.}~\bibnamefont {Amarni}}, \bibinfo {author} {\bibfnamefont
  {P.}~\bibnamefont {Astone}}, \bibinfo {author} {\bibfnamefont
  {G.}~\bibnamefont {Balestri}}, \bibinfo {author} {\bibfnamefont
  {G.}~\bibnamefont {Ballardin}}, \bibinfo {author} {\bibfnamefont
  {F.}~\bibnamefont {Barone}}, \bibinfo {author} {\bibfnamefont {J.-P.}\
  \bibnamefont {Baronick}}, \bibinfo {author} {\bibfnamefont {M.}~\bibnamefont
  {Barsuglia}}, \bibinfo {author} {\bibfnamefont {A.}~\bibnamefont {Basti}},
  \bibinfo {author} {\bibfnamefont {F.}~\bibnamefont {Basti}}, \bibinfo
  {author} {\bibfnamefont {T.~S.}\ \bibnamefont {Bauer}}, \bibinfo {author}
  {\bibfnamefont {V.}~\bibnamefont {Bavigadda}}, \bibinfo {author}
  {\bibfnamefont {M.}~\bibnamefont {Bejger}}, \bibinfo {author} {\bibfnamefont
  {M.~G.}\ \bibnamefont {Beker}}, \bibinfo {author} {\bibfnamefont
  {C.}~\bibnamefont {Belczynski}}, \bibinfo {author} {\bibfnamefont
  {D.}~\bibnamefont {Bersanetti}}, \bibinfo {author} {\bibfnamefont
  {A.}~\bibnamefont {Bertolini}}, \bibinfo {author} {\bibfnamefont
  {M.}~\bibnamefont {Bitossi}}, \bibinfo {author} {\bibfnamefont {M.~A.}\
  \bibnamefont {Bizouard}}, \bibinfo {author} {\bibfnamefont {S.}~\bibnamefont
  {Bloemen}}, \bibinfo {author} {\bibfnamefont {M.}~\bibnamefont {Blom}},
  \bibinfo {author} {\bibfnamefont {M.}~\bibnamefont {Boer}}, \bibinfo {author}
  {\bibfnamefont {G.}~\bibnamefont {Bogaert}}, \bibinfo {author} {\bibfnamefont
  {D.}~\bibnamefont {Bondi}}, \bibinfo {author} {\bibfnamefont
  {F.}~\bibnamefont {Bondu}}, \bibinfo {author} {\bibfnamefont
  {L.}~\bibnamefont {Bonelli}}, \bibinfo {author} {\bibfnamefont
  {R.}~\bibnamefont {Bonnand}}, \bibinfo {author} {\bibfnamefont
  {V.}~\bibnamefont {Boschi}}, \bibinfo {author} {\bibfnamefont
  {L.}~\bibnamefont {Bosi}}, \bibinfo {author} {\bibfnamefont {T.}~\bibnamefont
  {Bouedo}}, \bibinfo {author} {\bibfnamefont {C.}~\bibnamefont {Bradaschia}},
  \bibinfo {author} {\bibfnamefont {M.}~\bibnamefont {Branchesi}}, \bibinfo
  {author} {\bibfnamefont {T.}~\bibnamefont {Briant}}, \bibinfo {author}
  {\bibfnamefont {A.}~\bibnamefont {Brillet}}, \bibinfo {author} {\bibfnamefont
  {V.}~\bibnamefont {Brisson}}, \bibinfo {author} {\bibfnamefont
  {T.}~\bibnamefont {Bulik}}, \bibinfo {author} {\bibfnamefont {H.~J.}\
  \bibnamefont {Bulten}}, \bibinfo {author} {\bibfnamefont {D.}~\bibnamefont
  {Buskulic}}, \bibinfo {author} {\bibfnamefont {C.}~\bibnamefont {Buy}},
  \bibinfo {author} {\bibfnamefont {G.}~\bibnamefont {Cagnoli}}, \bibinfo
  {author} {\bibfnamefont {E.}~\bibnamefont {Calloni}}, \bibinfo {author}
  {\bibfnamefont {C.}~\bibnamefont {Campeggi}}, \bibinfo {author}
  {\bibfnamefont {B.}~\bibnamefont {Canuel}}, \bibinfo {author} {\bibfnamefont
  {F.}~\bibnamefont {Carbognani}}, \bibinfo {author} {\bibfnamefont
  {F.}~\bibnamefont {Cavalier}}, \bibinfo {author} {\bibfnamefont
  {R.}~\bibnamefont {Cavalieri}}, \ and\ \bibinfo {author} {\bibnamefont
  {et~al.}},\ }\href {\doibase 10.1088/0264-9381/32/2/024001} {\bibfield
  {journal} {\bibinfo  {journal} {Classical and Quantum Gravity}\ }\textbf
  {\bibinfo {volume} {32}},\ \bibinfo {pages} {024001} (\bibinfo {year}
  {2014})}\BibitemShut {NoStop}%
\bibitem [{\citenamefont {Akutsu}\ \emph {et~al.}(2019)\citenamefont {Akutsu}
  \emph {et~al.}}]{KAGRA}%
  \BibitemOpen
  \bibfield  {author} {\bibinfo {author} {\bibfnamefont {T.}~\bibnamefont
  {Akutsu}} \emph {et~al.} (\bibinfo {collaboration} {KAGRA}),\ }\href
  {\doibase 10.1038/s41550-018-0658-y} {\bibfield  {journal} {\bibinfo
  {journal} {Nature Astron.}\ }\textbf {\bibinfo {volume} {3}},\ \bibinfo
  {pages} {35} (\bibinfo {year} {2019})},\ \Eprint
  {http://arxiv.org/abs/1811.08079} {arXiv:1811.08079 [gr-qc]} \BibitemShut
  {NoStop}%
\bibitem [{\citenamefont {Collaboration}\ \emph
  {et~al.}(2021{\natexlab{a}})\citenamefont {Collaboration}, \citenamefont {the
  Virgo~Collaboration}, \citenamefont {the KAGRA~Collaboration}, \citenamefont
  {Abbott}, \citenamefont {Abbott}, \citenamefont {Acernese}, \citenamefont
  {Ackley}, \citenamefont {Adams}, \citenamefont {Adhikari}, \citenamefont
  {Adhikari}, \citenamefont {Adya}, \citenamefont {Affeldt}, \citenamefont
  {Agarwal}, \citenamefont {Agathos}, \citenamefont {Agatsuma}, \citenamefont
  {Aggarwal}, \citenamefont {Aguiar}, \citenamefont {Aiello}, \citenamefont
  {Ain},\ and\ \citenamefont {et~al.}}]{GWTC3}%
  \BibitemOpen
  \bibfield  {author} {\bibinfo {author} {\bibfnamefont {T.~L.~S.}\
  \bibnamefont {Collaboration}}, \bibinfo {author} {\bibnamefont {the
  Virgo~Collaboration}}, \bibinfo {author} {\bibnamefont {the
  KAGRA~Collaboration}}, \bibinfo {author} {\bibfnamefont {R.}~\bibnamefont
  {Abbott}}, \bibinfo {author} {\bibfnamefont {T.~D.}\ \bibnamefont {Abbott}},
  \bibinfo {author} {\bibfnamefont {F.}~\bibnamefont {Acernese}}, \bibinfo
  {author} {\bibfnamefont {K.}~\bibnamefont {Ackley}}, \bibinfo {author}
  {\bibfnamefont {C.}~\bibnamefont {Adams}}, \bibinfo {author} {\bibfnamefont
  {N.}~\bibnamefont {Adhikari}}, \bibinfo {author} {\bibfnamefont {R.~X.}\
  \bibnamefont {Adhikari}}, \bibinfo {author} {\bibfnamefont {V.~B.}\
  \bibnamefont {Adya}}, \bibinfo {author} {\bibfnamefont {C.}~\bibnamefont
  {Affeldt}}, \bibinfo {author} {\bibfnamefont {D.}~\bibnamefont {Agarwal}},
  \bibinfo {author} {\bibfnamefont {M.}~\bibnamefont {Agathos}}, \bibinfo
  {author} {\bibfnamefont {K.}~\bibnamefont {Agatsuma}}, \bibinfo {author}
  {\bibfnamefont {N.}~\bibnamefont {Aggarwal}}, \bibinfo {author}
  {\bibfnamefont {O.~D.}\ \bibnamefont {Aguiar}}, \bibinfo {author}
  {\bibfnamefont {L.}~\bibnamefont {Aiello}}, \bibinfo {author} {\bibfnamefont
  {A.}~\bibnamefont {Ain}}, \ and\ \bibinfo {author} {\bibnamefont {et~al.}},\
  }\href@noop {} {\  (\bibinfo {year} {2021}{\natexlab{a}})},\ \Eprint
  {http://arxiv.org/abs/2111.03606} {arXiv:2111.03606 [gr-qc]} \BibitemShut
  {NoStop}%
\bibitem [{\citenamefont {Abbott}\ \emph {et~al.}(2019)\citenamefont {Abbott},
  \citenamefont {Abbott}, \citenamefont {Abbott}, \citenamefont {Abraham},
  \citenamefont {Acernese}, \citenamefont {Ackley}, \citenamefont {Adams},
  \citenamefont {Adhikari}, \citenamefont {Adya}, \citenamefont {Affeldt},\
  and\ \citenamefont {et~al.}}]{GWTC1}%
  \BibitemOpen
  \bibfield  {author} {\bibinfo {author} {\bibfnamefont {B.}~\bibnamefont
  {Abbott}}, \bibinfo {author} {\bibfnamefont {R.}~\bibnamefont {Abbott}},
  \bibinfo {author} {\bibfnamefont {T.}~\bibnamefont {Abbott}}, \bibinfo
  {author} {\bibfnamefont {S.}~\bibnamefont {Abraham}}, \bibinfo {author}
  {\bibfnamefont {F.}~\bibnamefont {Acernese}}, \bibinfo {author}
  {\bibfnamefont {K.}~\bibnamefont {Ackley}}, \bibinfo {author} {\bibfnamefont
  {C.}~\bibnamefont {Adams}}, \bibinfo {author} {\bibfnamefont
  {R.}~\bibnamefont {Adhikari}}, \bibinfo {author} {\bibfnamefont
  {V.}~\bibnamefont {Adya}}, \bibinfo {author} {\bibfnamefont {C.}~\bibnamefont
  {Affeldt}}, \ and\ \bibinfo {author} {\bibnamefont {et~al.}},\ }\href
  {\doibase 10.1103/physrevx.9.031040} {\bibfield  {journal} {\bibinfo
  {journal} {Physical Review X}\ }\textbf {\bibinfo {volume} {9}} (\bibinfo
  {year} {2019}),\ 10.1103/physrevx.9.031040}\BibitemShut {NoStop}%
\bibitem [{\citenamefont {Abbott}\ \emph
  {et~al.}(2021{\natexlab{a}})\citenamefont {Abbott}, \citenamefont {Abbott},
  \citenamefont {Abraham}, \citenamefont {Acernese}, \citenamefont {Ackley},
  \citenamefont {Adams}, \citenamefont {Adams}, \citenamefont {Adhikari},
  \citenamefont {Adya}, \citenamefont {Affeldt},\ and\ \citenamefont
  {et~al.}}]{GWTC2}%
  \BibitemOpen
  \bibfield  {author} {\bibinfo {author} {\bibfnamefont {R.}~\bibnamefont
  {Abbott}}, \bibinfo {author} {\bibfnamefont {T.}~\bibnamefont {Abbott}},
  \bibinfo {author} {\bibfnamefont {S.}~\bibnamefont {Abraham}}, \bibinfo
  {author} {\bibfnamefont {F.}~\bibnamefont {Acernese}}, \bibinfo {author}
  {\bibfnamefont {K.}~\bibnamefont {Ackley}}, \bibinfo {author} {\bibfnamefont
  {A.}~\bibnamefont {Adams}}, \bibinfo {author} {\bibfnamefont
  {C.}~\bibnamefont {Adams}}, \bibinfo {author} {\bibfnamefont
  {R.}~\bibnamefont {Adhikari}}, \bibinfo {author} {\bibfnamefont
  {V.}~\bibnamefont {Adya}}, \bibinfo {author} {\bibfnamefont {C.}~\bibnamefont
  {Affeldt}}, \ and\ \bibinfo {author} {\bibnamefont {et~al.}},\ }\href
  {\doibase 10.1103/physrevx.11.021053} {\bibfield  {journal} {\bibinfo
  {journal} {Physical Review X}\ }\textbf {\bibinfo {volume} {11}} (\bibinfo
  {year} {2021}{\natexlab{a}}),\ 10.1103/physrevx.11.021053}\BibitemShut
  {NoStop}%
\bibitem [{\citenamefont {Collaboration}\ \emph
  {et~al.}(2021{\natexlab{b}})\citenamefont {Collaboration}, \citenamefont {the
  Virgo~Collaboration}, \citenamefont {the KAGRA~Collaboration}, \citenamefont
  {Abbott}, \citenamefont {Abbott}, \citenamefont {Acernese}, \citenamefont
  {Ackley}, \citenamefont {Adams}, \citenamefont {Adhikari}, \citenamefont
  {Adhikari}, \citenamefont {Adya}, \citenamefont {Affeldt}, \citenamefont
  {Agarwal}, \citenamefont {Agathos}, \citenamefont {Agatsuma}, \citenamefont
  {Aggarwal}, \citenamefont {Aguiar}, \citenamefont {Aiello}, \citenamefont
  {Ain}, \citenamefont {Ajith}, \citenamefont {Akutsu}, \citenamefont
  {Albanesi}, \citenamefont {Allocca}, \citenamefont {Altin}, \citenamefont
  {Amato}, \citenamefont {Anand}, \citenamefont {Anand}, \citenamefont
  {Ananyeva}, \citenamefont {Anderson}, \citenamefont {Anderson}, \citenamefont
  {Ando}, \citenamefont {Andrade}, \citenamefont {Andres}, \citenamefont
  {Andrić}, \citenamefont {Angelova}, \citenamefont {Ansoldi},\ and\
  \citenamefont {et~al.}}]{popgwtc3}%
  \BibitemOpen
  \bibfield  {author} {\bibinfo {author} {\bibfnamefont {T.~L.~S.}\
  \bibnamefont {Collaboration}}, \bibinfo {author} {\bibnamefont {the
  Virgo~Collaboration}}, \bibinfo {author} {\bibnamefont {the
  KAGRA~Collaboration}}, \bibinfo {author} {\bibfnamefont {R.}~\bibnamefont
  {Abbott}}, \bibinfo {author} {\bibfnamefont {T.~D.}\ \bibnamefont {Abbott}},
  \bibinfo {author} {\bibfnamefont {F.}~\bibnamefont {Acernese}}, \bibinfo
  {author} {\bibfnamefont {K.}~\bibnamefont {Ackley}}, \bibinfo {author}
  {\bibfnamefont {C.}~\bibnamefont {Adams}}, \bibinfo {author} {\bibfnamefont
  {N.}~\bibnamefont {Adhikari}}, \bibinfo {author} {\bibfnamefont {R.~X.}\
  \bibnamefont {Adhikari}}, \bibinfo {author} {\bibfnamefont {V.~B.}\
  \bibnamefont {Adya}}, \bibinfo {author} {\bibfnamefont {C.}~\bibnamefont
  {Affeldt}}, \bibinfo {author} {\bibfnamefont {D.}~\bibnamefont {Agarwal}},
  \bibinfo {author} {\bibfnamefont {M.}~\bibnamefont {Agathos}}, \bibinfo
  {author} {\bibfnamefont {K.}~\bibnamefont {Agatsuma}}, \bibinfo {author}
  {\bibfnamefont {N.}~\bibnamefont {Aggarwal}}, \bibinfo {author}
  {\bibfnamefont {O.~D.}\ \bibnamefont {Aguiar}}, \bibinfo {author}
  {\bibfnamefont {L.}~\bibnamefont {Aiello}}, \bibinfo {author} {\bibfnamefont
  {A.}~\bibnamefont {Ain}}, \bibinfo {author} {\bibfnamefont {P.}~\bibnamefont
  {Ajith}}, \bibinfo {author} {\bibfnamefont {T.}~\bibnamefont {Akutsu}},
  \bibinfo {author} {\bibfnamefont {S.}~\bibnamefont {Albanesi}}, \bibinfo
  {author} {\bibfnamefont {A.}~\bibnamefont {Allocca}}, \bibinfo {author}
  {\bibfnamefont {P.~A.}\ \bibnamefont {Altin}}, \bibinfo {author}
  {\bibfnamefont {A.}~\bibnamefont {Amato}}, \bibinfo {author} {\bibfnamefont
  {C.}~\bibnamefont {Anand}}, \bibinfo {author} {\bibfnamefont
  {S.}~\bibnamefont {Anand}}, \bibinfo {author} {\bibfnamefont
  {A.}~\bibnamefont {Ananyeva}}, \bibinfo {author} {\bibfnamefont {S.~B.}\
  \bibnamefont {Anderson}}, \bibinfo {author} {\bibfnamefont {W.~G.}\
  \bibnamefont {Anderson}}, \bibinfo {author} {\bibfnamefont {M.}~\bibnamefont
  {Ando}}, \bibinfo {author} {\bibfnamefont {T.}~\bibnamefont {Andrade}},
  \bibinfo {author} {\bibfnamefont {N.}~\bibnamefont {Andres}}, \bibinfo
  {author} {\bibfnamefont {T.}~\bibnamefont {Andrić}}, \bibinfo {author}
  {\bibfnamefont {S.~V.}\ \bibnamefont {Angelova}}, \bibinfo {author}
  {\bibfnamefont {S.}~\bibnamefont {Ansoldi}}, \ and\ \bibinfo {author}
  {\bibnamefont {et~al.}},\ }\href@noop {} {\enquote {\bibinfo {title} {The
  population of merging compact binaries inferred using gravitational waves
  through gwtc-3},}\ } (\bibinfo {year} {2021}{\natexlab{b}}),\ \Eprint
  {http://arxiv.org/abs/2111.03634} {arXiv:2111.03634 [astro-ph.HE]}
  \BibitemShut {NoStop}%
\bibitem [{\citenamefont {Maggiore}\ \emph {et~al.}(2020)\citenamefont
  {Maggiore}, \citenamefont {Broeck}, \citenamefont {Bartolo}, \citenamefont
  {Belgacem}, \citenamefont {Bertacca}, \citenamefont {Bizouard}, \citenamefont
  {Branchesi}, \citenamefont {Clesse}, \citenamefont {Foffa}, \citenamefont
  {García-Bellido},\ and\ \citenamefont {et~al.}}]{EinsteinTelescope}%
  \BibitemOpen
  \bibfield  {author} {\bibinfo {author} {\bibfnamefont {M.}~\bibnamefont
  {Maggiore}}, \bibinfo {author} {\bibfnamefont {C.~V.~D.}\ \bibnamefont
  {Broeck}}, \bibinfo {author} {\bibfnamefont {N.}~\bibnamefont {Bartolo}},
  \bibinfo {author} {\bibfnamefont {E.}~\bibnamefont {Belgacem}}, \bibinfo
  {author} {\bibfnamefont {D.}~\bibnamefont {Bertacca}}, \bibinfo {author}
  {\bibfnamefont {M.~A.}\ \bibnamefont {Bizouard}}, \bibinfo {author}
  {\bibfnamefont {M.}~\bibnamefont {Branchesi}}, \bibinfo {author}
  {\bibfnamefont {S.}~\bibnamefont {Clesse}}, \bibinfo {author} {\bibfnamefont
  {S.}~\bibnamefont {Foffa}}, \bibinfo {author} {\bibfnamefont
  {J.}~\bibnamefont {García-Bellido}}, \ and\ \bibinfo {author} {\bibnamefont
  {et~al.}},\ }\href {\doibase 10.1088/1475-7516/2020/03/050} {\bibfield
  {journal} {\bibinfo  {journal} {Journal of Cosmology and Astroparticle
  Physics}\ }\textbf {\bibinfo {volume} {2020}},\ \bibinfo {pages} {050–050}
  (\bibinfo {year} {2020})}\BibitemShut {NoStop}%
\bibitem [{\citenamefont {Reitze}\ \emph {et~al.}(2019)\citenamefont {Reitze},
  \citenamefont {Adhikari}, \citenamefont {Ballmer}, \citenamefont {Barish},
  \citenamefont {Barsotti}, \citenamefont {Billingsley}, \citenamefont {Brown},
  \citenamefont {Chen}, \citenamefont {Coyne}, \citenamefont {Eisenstein},
  \citenamefont {Evans}, \citenamefont {Fritschel}, \citenamefont {Hall},
  \citenamefont {Lazzarini}, \citenamefont {Lovelace}, \citenamefont {Read},
  \citenamefont {Sathyaprakash}, \citenamefont {Shoemaker}, \citenamefont
  {Smith}, \citenamefont {Torrie}, \citenamefont {Vitale}, \citenamefont
  {Weiss}, \citenamefont {Wipf},\ and\ \citenamefont
  {Zucker}}]{CosmicExplorer}%
  \BibitemOpen
  \bibfield  {author} {\bibinfo {author} {\bibfnamefont {D.}~\bibnamefont
  {Reitze}}, \bibinfo {author} {\bibfnamefont {R.~X.}\ \bibnamefont
  {Adhikari}}, \bibinfo {author} {\bibfnamefont {S.}~\bibnamefont {Ballmer}},
  \bibinfo {author} {\bibfnamefont {B.}~\bibnamefont {Barish}}, \bibinfo
  {author} {\bibfnamefont {L.}~\bibnamefont {Barsotti}}, \bibinfo {author}
  {\bibfnamefont {G.}~\bibnamefont {Billingsley}}, \bibinfo {author}
  {\bibfnamefont {D.~A.}\ \bibnamefont {Brown}}, \bibinfo {author}
  {\bibfnamefont {Y.}~\bibnamefont {Chen}}, \bibinfo {author} {\bibfnamefont
  {D.}~\bibnamefont {Coyne}}, \bibinfo {author} {\bibfnamefont
  {R.}~\bibnamefont {Eisenstein}}, \bibinfo {author} {\bibfnamefont
  {M.}~\bibnamefont {Evans}}, \bibinfo {author} {\bibfnamefont
  {P.}~\bibnamefont {Fritschel}}, \bibinfo {author} {\bibfnamefont {E.~D.}\
  \bibnamefont {Hall}}, \bibinfo {author} {\bibfnamefont {A.}~\bibnamefont
  {Lazzarini}}, \bibinfo {author} {\bibfnamefont {G.}~\bibnamefont {Lovelace}},
  \bibinfo {author} {\bibfnamefont {J.}~\bibnamefont {Read}}, \bibinfo {author}
  {\bibfnamefont {B.~S.}\ \bibnamefont {Sathyaprakash}}, \bibinfo {author}
  {\bibfnamefont {D.}~\bibnamefont {Shoemaker}}, \bibinfo {author}
  {\bibfnamefont {J.}~\bibnamefont {Smith}}, \bibinfo {author} {\bibfnamefont
  {C.}~\bibnamefont {Torrie}}, \bibinfo {author} {\bibfnamefont
  {S.}~\bibnamefont {Vitale}}, \bibinfo {author} {\bibfnamefont
  {R.}~\bibnamefont {Weiss}}, \bibinfo {author} {\bibfnamefont
  {C.}~\bibnamefont {Wipf}}, \ and\ \bibinfo {author} {\bibfnamefont
  {M.}~\bibnamefont {Zucker}},\ }\href
  {https://baas.aas.org/pub/2020n7i035/release/1} {\enquote {\bibinfo {title}
  {Cosmic explorer: The u.s. contribution to gravitational-wave astronomy
  beyond ligo},}\ } (\bibinfo {year} {2019})\BibitemShut {NoStop}%
\bibitem [{\citenamefont {Abbott}\ \emph {et~al.}(2020)\citenamefont {Abbott},
  \citenamefont {Abbott}, \citenamefont {Abbott}, \citenamefont {Abraham},
  \citenamefont {Acernese}, \citenamefont {Ackley}, \citenamefont {Adams},
  \citenamefont {Adya}, \citenamefont {Affeldt}, \citenamefont {Agathos},\ and\
  \citenamefont {et~al.}}]{prospects_LVK}%
  \BibitemOpen
  \bibfield  {author} {\bibinfo {author} {\bibfnamefont {B.~P.}\ \bibnamefont
  {Abbott}}, \bibinfo {author} {\bibfnamefont {R.}~\bibnamefont {Abbott}},
  \bibinfo {author} {\bibfnamefont {T.~D.}\ \bibnamefont {Abbott}}, \bibinfo
  {author} {\bibfnamefont {S.}~\bibnamefont {Abraham}}, \bibinfo {author}
  {\bibfnamefont {F.}~\bibnamefont {Acernese}}, \bibinfo {author}
  {\bibfnamefont {K.}~\bibnamefont {Ackley}}, \bibinfo {author} {\bibfnamefont
  {C.}~\bibnamefont {Adams}}, \bibinfo {author} {\bibfnamefont {V.~B.}\
  \bibnamefont {Adya}}, \bibinfo {author} {\bibfnamefont {C.}~\bibnamefont
  {Affeldt}}, \bibinfo {author} {\bibfnamefont {M.}~\bibnamefont {Agathos}}, \
  and\ \bibinfo {author} {\bibnamefont {et~al.}},\ }\href {\doibase
  10.1007/s41114-020-00026-9} {\bibfield  {journal} {\bibinfo  {journal}
  {Living Reviews in Relativity}\ }\textbf {\bibinfo {volume} {23}} (\bibinfo
  {year} {2020}),\ 10.1007/s41114-020-00026-9}\BibitemShut {NoStop}%
\bibitem [{\citenamefont {Perkins}\ \emph {et~al.}(2021)\citenamefont
  {Perkins}, \citenamefont {Yunes},\ and\ \citenamefont
  {Berti}}]{prob_population_fundamental_physics}%
  \BibitemOpen
  \bibfield  {author} {\bibinfo {author} {\bibfnamefont {S.~E.}\ \bibnamefont
  {Perkins}}, \bibinfo {author} {\bibfnamefont {N.}~\bibnamefont {Yunes}}, \
  and\ \bibinfo {author} {\bibfnamefont {E.}~\bibnamefont {Berti}},\ }\href
  {\doibase 10.1103/PhysRevD.103.044024} {\bibfield  {journal} {\bibinfo
  {journal} {Phys. Rev. D}\ }\textbf {\bibinfo {volume} {103}},\ \bibinfo
  {pages} {044024} (\bibinfo {year} {2021})}\BibitemShut {NoStop}%
\bibitem [{\citenamefont {Wysocki}\ \emph {et~al.}(2018)\citenamefont
  {Wysocki}, \citenamefont {Gerosa}, \citenamefont {O'Shaughnessy},
  \citenamefont {Belczynski}, \citenamefont {Gladysz}, \citenamefont {Berti},
  \citenamefont {Kesden},\ and\ \citenamefont {Holz}}]{PhysRevD.97.043014}%
  \BibitemOpen
  \bibfield  {author} {\bibinfo {author} {\bibfnamefont {D.}~\bibnamefont
  {Wysocki}}, \bibinfo {author} {\bibfnamefont {D.}~\bibnamefont {Gerosa}},
  \bibinfo {author} {\bibfnamefont {R.}~\bibnamefont {O'Shaughnessy}}, \bibinfo
  {author} {\bibfnamefont {K.}~\bibnamefont {Belczynski}}, \bibinfo {author}
  {\bibfnamefont {W.}~\bibnamefont {Gladysz}}, \bibinfo {author} {\bibfnamefont
  {E.}~\bibnamefont {Berti}}, \bibinfo {author} {\bibfnamefont
  {M.}~\bibnamefont {Kesden}}, \ and\ \bibinfo {author} {\bibfnamefont {D.~E.}\
  \bibnamefont {Holz}},\ }\href {\doibase 10.1103/PhysRevD.97.043014}
  {\bibfield  {journal} {\bibinfo  {journal} {Phys. Rev. D}\ }\textbf {\bibinfo
  {volume} {97}},\ \bibinfo {pages} {043014} (\bibinfo {year}
  {2018})}\BibitemShut {NoStop}%
\bibitem [{\citenamefont {Wysocki}\ \emph {et~al.}(2019)\citenamefont
  {Wysocki}, \citenamefont {Lange},\ and\ \citenamefont
  {O'Shaughnessy}}]{PhysRevD.100.043012}%
  \BibitemOpen
  \bibfield  {author} {\bibinfo {author} {\bibfnamefont {D.}~\bibnamefont
  {Wysocki}}, \bibinfo {author} {\bibfnamefont {J.}~\bibnamefont {Lange}}, \
  and\ \bibinfo {author} {\bibfnamefont {R.}~\bibnamefont {O'Shaughnessy}},\
  }\href {\doibase 10.1103/PhysRevD.100.043012} {\bibfield  {journal} {\bibinfo
   {journal} {Phys. Rev. D}\ }\textbf {\bibinfo {volume} {100}},\ \bibinfo
  {pages} {043012} (\bibinfo {year} {2019})}\BibitemShut {NoStop}%
\bibitem [{\citenamefont {Wiktorowicz}\ \emph {et~al.}(2019)\citenamefont
  {Wiktorowicz}, \citenamefont {Wyrzykowski}, \citenamefont {Chruslinska},
  \citenamefont {Klencki}, \citenamefont {Rybicki},\ and\ \citenamefont
  {Belczynski}}]{stellarmass_natalkick}%
  \BibitemOpen
  \bibfield  {author} {\bibinfo {author} {\bibfnamefont {G.}~\bibnamefont
  {Wiktorowicz}}, \bibinfo {author} {\bibfnamefont {{\L}.}~\bibnamefont
  {Wyrzykowski}}, \bibinfo {author} {\bibfnamefont {M.}~\bibnamefont
  {Chruslinska}}, \bibinfo {author} {\bibfnamefont {J.}~\bibnamefont
  {Klencki}}, \bibinfo {author} {\bibfnamefont {K.~A.}\ \bibnamefont
  {Rybicki}}, \ and\ \bibinfo {author} {\bibfnamefont {K.}~\bibnamefont
  {Belczynski}},\ }\href {\doibase 10.3847/1538-4357/ab45e6} {\bibfield
  {journal} {\bibinfo  {journal} {The Astrophysical Journal}\ }\textbf
  {\bibinfo {volume} {885}},\ \bibinfo {pages} {1} (\bibinfo {year}
  {2019})}\BibitemShut {NoStop}%
\bibitem [{\citenamefont {Barrett}\ \emph {et~al.}(2016)\citenamefont
  {Barrett}, \citenamefont {Mandel}, \citenamefont {Neijssel}, \citenamefont
  {Stevenson},\ and\ \citenamefont {Vigna-Gómez}}]{m1}%
  \BibitemOpen
  \bibfield  {author} {\bibinfo {author} {\bibfnamefont {J.~W.}\ \bibnamefont
  {Barrett}}, \bibinfo {author} {\bibfnamefont {I.}~\bibnamefont {Mandel}},
  \bibinfo {author} {\bibfnamefont {C.~J.}\ \bibnamefont {Neijssel}}, \bibinfo
  {author} {\bibfnamefont {S.}~\bibnamefont {Stevenson}}, \ and\ \bibinfo
  {author} {\bibfnamefont {A.}~\bibnamefont {Vigna-Gómez}},\ }\href {\doibase
  10.1017/s1743921317000059} {\bibfield  {journal} {\bibinfo  {journal}
  {Proceedings of the International Astronomical Union}\ }\textbf {\bibinfo
  {volume} {12}},\ \bibinfo {pages} {46–50} (\bibinfo {year}
  {2016})}\BibitemShut {NoStop}%
\bibitem [{\citenamefont {Belczynski}\ \emph {et~al.}(2010)\citenamefont
  {Belczynski}, \citenamefont {Dominik}, \citenamefont {Bulik}, \citenamefont
  {O'Shaughnessy}, \citenamefont {Fryer},\ and\ \citenamefont
  {Holz}}]{metallicity}%
  \BibitemOpen
  \bibfield  {author} {\bibinfo {author} {\bibfnamefont {K.}~\bibnamefont
  {Belczynski}}, \bibinfo {author} {\bibfnamefont {M.}~\bibnamefont {Dominik}},
  \bibinfo {author} {\bibfnamefont {T.}~\bibnamefont {Bulik}}, \bibinfo
  {author} {\bibfnamefont {R.}~\bibnamefont {O'Shaughnessy}}, \bibinfo {author}
  {\bibfnamefont {C.}~\bibnamefont {Fryer}}, \ and\ \bibinfo {author}
  {\bibfnamefont {D.~E.}\ \bibnamefont {Holz}},\ }\href {\doibase
  10.1088/2041-8205/715/2/l138} {\ \textbf {\bibinfo {volume} {715}},\ \bibinfo
  {pages} {L138} (\bibinfo {year} {2010})}\BibitemShut {NoStop}%
\bibitem [{\citenamefont {Taylor}\ and\ \citenamefont
  {Gerosa}(2018)}]{taylor2018mining}%
  \BibitemOpen
  \bibfield  {author} {\bibinfo {author} {\bibfnamefont {S.~R.}\ \bibnamefont
  {Taylor}}\ and\ \bibinfo {author} {\bibfnamefont {D.}~\bibnamefont
  {Gerosa}},\ }\href@noop {} {\bibfield  {journal} {\bibinfo  {journal}
  {Physical Review D}\ }\textbf {\bibinfo {volume} {98}},\ \bibinfo {pages}
  {083017} (\bibinfo {year} {2018})}\BibitemShut {NoStop}%
\bibitem [{\citenamefont {Wong}\ and\ \citenamefont
  {Gerosa}(2019)}]{PhysRevD.100.083015}%
  \BibitemOpen
  \bibfield  {author} {\bibinfo {author} {\bibfnamefont {K.~W.~K.}\
  \bibnamefont {Wong}}\ and\ \bibinfo {author} {\bibfnamefont {D.}~\bibnamefont
  {Gerosa}},\ }\href {\doibase 10.1103/PhysRevD.100.083015} {\bibfield
  {journal} {\bibinfo  {journal} {Phys. Rev. D}\ }\textbf {\bibinfo {volume}
  {100}},\ \bibinfo {pages} {083015} (\bibinfo {year} {2019})}\BibitemShut
  {NoStop}%
\bibitem [{\citenamefont {Wong}\ \emph {et~al.}(2020)\citenamefont {Wong},
  \citenamefont {Contardo},\ and\ \citenamefont {Ho}}]{Wong:2020jdt}%
  \BibitemOpen
  \bibfield  {author} {\bibinfo {author} {\bibfnamefont {K.~W.~K.}\
  \bibnamefont {Wong}}, \bibinfo {author} {\bibfnamefont {G.}~\bibnamefont
  {Contardo}}, \ and\ \bibinfo {author} {\bibfnamefont {S.}~\bibnamefont
  {Ho}},\ }\href {\doibase 10.1103/PhysRevD.101.123005} {\bibfield  {journal}
  {\bibinfo  {journal} {Phys. Rev. D}\ }\textbf {\bibinfo {volume} {101}},\
  \bibinfo {pages} {123005} (\bibinfo {year} {2020})},\ \Eprint
  {http://arxiv.org/abs/2002.09491} {arXiv:2002.09491 [astro-ph.IM]}
  \BibitemShut {NoStop}%
\bibitem [{\citenamefont {Abbott}\ \emph
  {et~al.}(2021{\natexlab{b}})\citenamefont {Abbott}, \citenamefont {Abbott},
  \citenamefont {Abraham}, \citenamefont {Acernese}, \citenamefont {Ackley},
  \citenamefont {Adams}, \citenamefont {Adams}, \citenamefont {Adhikari},
  \citenamefont {Adya}, \citenamefont {Affeldt},\ and\ \citenamefont
  {et~al.}}]{POPgwtc2}%
  \BibitemOpen
  \bibfield  {author} {\bibinfo {author} {\bibfnamefont {R.}~\bibnamefont
  {Abbott}}, \bibinfo {author} {\bibfnamefont {T.~D.}\ \bibnamefont {Abbott}},
  \bibinfo {author} {\bibfnamefont {S.}~\bibnamefont {Abraham}}, \bibinfo
  {author} {\bibfnamefont {F.}~\bibnamefont {Acernese}}, \bibinfo {author}
  {\bibfnamefont {K.}~\bibnamefont {Ackley}}, \bibinfo {author} {\bibfnamefont
  {A.}~\bibnamefont {Adams}}, \bibinfo {author} {\bibfnamefont
  {C.}~\bibnamefont {Adams}}, \bibinfo {author} {\bibfnamefont {R.~X.}\
  \bibnamefont {Adhikari}}, \bibinfo {author} {\bibfnamefont {V.~B.}\
  \bibnamefont {Adya}}, \bibinfo {author} {\bibfnamefont {C.}~\bibnamefont
  {Affeldt}}, \ and\ \bibinfo {author} {\bibnamefont {et~al.}},\ }\href
  {\doibase 10.3847/2041-8213/abe949} {\bibfield  {journal} {\bibinfo
  {journal} {The Astrophysical Journal Letters}\ }\textbf {\bibinfo {volume}
  {913}},\ \bibinfo {pages} {L7} (\bibinfo {year}
  {2021}{\natexlab{b}})}\BibitemShut {NoStop}%
\bibitem [{\citenamefont {Chen}\ \emph {et~al.}(2017)\citenamefont {Chen},
  \citenamefont {Essick}, \citenamefont {Vitale}, \citenamefont {Holz},\ and\
  \citenamefont {Katsavounidis}}]{instrubias}%
  \BibitemOpen
  \bibfield  {author} {\bibinfo {author} {\bibfnamefont {H.-Y.}\ \bibnamefont
  {Chen}}, \bibinfo {author} {\bibfnamefont {R.}~\bibnamefont {Essick}},
  \bibinfo {author} {\bibfnamefont {S.}~\bibnamefont {Vitale}}, \bibinfo
  {author} {\bibfnamefont {D.~E.}\ \bibnamefont {Holz}}, \ and\ \bibinfo
  {author} {\bibfnamefont {E.}~\bibnamefont {Katsavounidis}},\ }\href {\doibase
  10.3847/1538-4357/835/1/31} {\bibfield  {journal} {\bibinfo  {journal} {The
  Astrophysical Journal}\ }\textbf {\bibinfo {volume} {835}},\ \bibinfo {pages}
  {31} (\bibinfo {year} {2017})}\BibitemShut {NoStop}%
\bibitem [{\citenamefont {{Vitale}}\ \emph {et~al.}(2020)\citenamefont
  {{Vitale}}, \citenamefont {{Gerosa}}, \citenamefont {{Farr}},\ and\
  \citenamefont {{Taylor}}}]{selectionbias_pop_inference}%
  \BibitemOpen
  \bibfield  {author} {\bibinfo {author} {\bibfnamefont {S.}~\bibnamefont
  {{Vitale}}}, \bibinfo {author} {\bibfnamefont {D.}~\bibnamefont {{Gerosa}}},
  \bibinfo {author} {\bibfnamefont {W.~M.}\ \bibnamefont {{Farr}}}, \ and\
  \bibinfo {author} {\bibfnamefont {S.~R.}\ \bibnamefont {{Taylor}}},\
  }\href@noop {} {\bibfield  {journal} {\bibinfo  {journal} {arXiv e-prints}\
  ,\ \bibinfo {eid} {arXiv:2007.05579}} (\bibinfo {year} {2020})},\ \Eprint
  {http://arxiv.org/abs/2007.05579} {arXiv:2007.05579 [astro-ph.IM]}
  \BibitemShut {NoStop}%
\bibitem [{\citenamefont {Thrane}\ and\ \citenamefont
  {Talbot}(2019)}]{para_estimation}%
  \BibitemOpen
  \bibfield  {author} {\bibinfo {author} {\bibfnamefont {E.}~\bibnamefont
  {Thrane}}\ and\ \bibinfo {author} {\bibfnamefont {C.}~\bibnamefont
  {Talbot}},\ }\href {\doibase 10.1017/pasa.2019.2} {\bibfield  {journal}
  {\bibinfo  {journal} {Publications of the Astronomical Society of Australia}\
  }\textbf {\bibinfo {volume} {36}} (\bibinfo {year} {2019}),\
  10.1017/pasa.2019.2}\BibitemShut {NoStop}%
\bibitem [{\citenamefont {Vitale}\ \emph {et~al.}(2017)\citenamefont {Vitale},
  \citenamefont {Gerosa}, \citenamefont {Haster}, \citenamefont
  {Chatziioannou},\ and\ \citenamefont {Zimmerman}}]{PhysRevLett.119.251103}%
  \BibitemOpen
  \bibfield  {author} {\bibinfo {author} {\bibfnamefont {S.}~\bibnamefont
  {Vitale}}, \bibinfo {author} {\bibfnamefont {D.}~\bibnamefont {Gerosa}},
  \bibinfo {author} {\bibfnamefont {C.-J.}\ \bibnamefont {Haster}}, \bibinfo
  {author} {\bibfnamefont {K.}~\bibnamefont {Chatziioannou}}, \ and\ \bibinfo
  {author} {\bibfnamefont {A.}~\bibnamefont {Zimmerman}},\ }\href {\doibase
  10.1103/PhysRevLett.119.251103} {\bibfield  {journal} {\bibinfo  {journal}
  {Phys. Rev. Lett.}\ }\textbf {\bibinfo {volume} {119}},\ \bibinfo {pages}
  {251103} (\bibinfo {year} {2017})}\BibitemShut {NoStop}%
\bibitem [{\citenamefont {Pankow}\ \emph {et~al.}(2017)\citenamefont {Pankow},
  \citenamefont {Sampson}, \citenamefont {Perri}, \citenamefont {Chase},
  \citenamefont {Coughlin}, \citenamefont {Zevin},\ and\ \citenamefont
  {Kalogera}}]{prior}%
  \BibitemOpen
  \bibfield  {author} {\bibinfo {author} {\bibfnamefont {C.}~\bibnamefont
  {Pankow}}, \bibinfo {author} {\bibfnamefont {L.}~\bibnamefont {Sampson}},
  \bibinfo {author} {\bibfnamefont {L.}~\bibnamefont {Perri}}, \bibinfo
  {author} {\bibfnamefont {E.}~\bibnamefont {Chase}}, \bibinfo {author}
  {\bibfnamefont {S.}~\bibnamefont {Coughlin}}, \bibinfo {author}
  {\bibfnamefont {M.}~\bibnamefont {Zevin}}, \ and\ \bibinfo {author}
  {\bibfnamefont {V.}~\bibnamefont {Kalogera}},\ }\href {\doibase
  10.3847/1538-4357/834/2/154} {\bibfield  {journal} {\bibinfo  {journal} {The
  Astrophysical Journal}\ }\textbf {\bibinfo {volume} {834}},\ \bibinfo {pages}
  {154} (\bibinfo {year} {2017})}\BibitemShut {NoStop}%
\bibitem [{\citenamefont {Hogg}\ \emph {et~al.}(2010)\citenamefont {Hogg},
  \citenamefont {Myers},\ and\ \citenamefont {Bovy}}]{DavidHogg_HBA}%
  \BibitemOpen
  \bibfield  {author} {\bibinfo {author} {\bibfnamefont {D.~W.}\ \bibnamefont
  {Hogg}}, \bibinfo {author} {\bibfnamefont {A.~D.}\ \bibnamefont {Myers}}, \
  and\ \bibinfo {author} {\bibfnamefont {J.}~\bibnamefont {Bovy}},\ }\href
  {\doibase 10.1088/0004-637x/725/2/2166} {\bibfield  {journal} {\bibinfo
  {journal} {The Astrophysical Journal}\ }\textbf {\bibinfo {volume} {725}},\
  \bibinfo {pages} {2166–2175} (\bibinfo {year} {2010})}\BibitemShut
  {NoStop}%
\bibitem [{\citenamefont {Mandel}\ \emph {et~al.}(2019)\citenamefont {Mandel},
  \citenamefont {Farr},\ and\ \citenamefont {Gair}}]{stat_frame_sb}%
  \BibitemOpen
  \bibfield  {author} {\bibinfo {author} {\bibfnamefont {I.}~\bibnamefont
  {Mandel}}, \bibinfo {author} {\bibfnamefont {W.~M.}\ \bibnamefont {Farr}}, \
  and\ \bibinfo {author} {\bibfnamefont {J.~R.}\ \bibnamefont {Gair}},\ }\href
  {\doibase 10.1093/mnras/stz896} {\bibfield  {journal} {\bibinfo  {journal}
  {Monthly Notices of the Royal Astronomical Society}\ }\textbf {\bibinfo
  {volume} {486}},\ \bibinfo {pages} {1086–1093} (\bibinfo {year}
  {2019})}\BibitemShut {NoStop}%
\bibitem [{\citenamefont {Talbot}\ and\ \citenamefont {Thrane}(2018)}]{GWHBA}%
  \BibitemOpen
  \bibfield  {author} {\bibinfo {author} {\bibfnamefont {C.}~\bibnamefont
  {Talbot}}\ and\ \bibinfo {author} {\bibfnamefont {E.}~\bibnamefont
  {Thrane}},\ }\href {\doibase 10.3847/1538-4357/aab34c} {\ \textbf {\bibinfo
  {volume} {856}},\ \bibinfo {pages} {173} (\bibinfo {year}
  {2018})}\BibitemShut {NoStop}%
\bibitem [{\citenamefont {Himemoto}\ \emph {et~al.}(2021)\citenamefont
  {Himemoto}, \citenamefont {Nishizawa},\ and\ \citenamefont
  {Taruya}}]{3rdoverlap_signal}%
  \BibitemOpen
  \bibfield  {author} {\bibinfo {author} {\bibfnamefont {Y.}~\bibnamefont
  {Himemoto}}, \bibinfo {author} {\bibfnamefont {A.}~\bibnamefont {Nishizawa}},
  \ and\ \bibinfo {author} {\bibfnamefont {A.}~\bibnamefont {Taruya}},\ }\href
  {\doibase 10.1103/PhysRevD.104.044010} {\bibfield  {journal} {\bibinfo
  {journal} {Phys. Rev. D}\ }\textbf {\bibinfo {volume} {104}},\ \bibinfo
  {pages} {044010} (\bibinfo {year} {2021})}\BibitemShut {NoStop}%
\bibitem [{\citenamefont {Martynov}\ \emph {et~al.}(2016)\citenamefont
  {Martynov}, \citenamefont {Hall}, \citenamefont {Abbott}, \citenamefont
  {Abbott}, \citenamefont {Abbott}, \citenamefont {Adams}, \citenamefont
  {Adhikari}, \citenamefont {Anderson}, \citenamefont {Anderson}, \citenamefont
  {Arai},\ and\ \citenamefont {et~al.}}]{sensitivity_ligovirgo}%
  \BibitemOpen
  \bibfield  {author} {\bibinfo {author} {\bibfnamefont {D.}~\bibnamefont
  {Martynov}}, \bibinfo {author} {\bibfnamefont {E.}~\bibnamefont {Hall}},
  \bibinfo {author} {\bibfnamefont {B.}~\bibnamefont {Abbott}}, \bibinfo
  {author} {\bibfnamefont {R.}~\bibnamefont {Abbott}}, \bibinfo {author}
  {\bibfnamefont {T.}~\bibnamefont {Abbott}}, \bibinfo {author} {\bibfnamefont
  {C.}~\bibnamefont {Adams}}, \bibinfo {author} {\bibfnamefont
  {R.}~\bibnamefont {Adhikari}}, \bibinfo {author} {\bibfnamefont
  {R.}~\bibnamefont {Anderson}}, \bibinfo {author} {\bibfnamefont
  {S.}~\bibnamefont {Anderson}}, \bibinfo {author} {\bibfnamefont
  {K.}~\bibnamefont {Arai}}, \ and\ \bibinfo {author} {\bibnamefont {et~al.}},\
  }\href {\doibase 10.1103/physrevd.93.112004} {\bibfield  {journal} {\bibinfo
  {journal} {Physical Review D}\ }\textbf {\bibinfo {volume} {93}} (\bibinfo
  {year} {2016}),\ 10.1103/physrevd.93.112004}\BibitemShut {NoStop}%
\bibitem [{\citenamefont {Singer}\ \emph {et~al.}(2016)\citenamefont {Singer},
  \citenamefont {Chen}, \citenamefont {Holz}, \citenamefont {Farr},
  \citenamefont {Price}, \citenamefont {Raymond}, \citenamefont {Cenko},
  \citenamefont {Gehrels}, \citenamefont {Cannizzo}, \citenamefont {Kasliwal},\
  and\ \citenamefont {et~al.}}]{sensitivity_1}%
  \BibitemOpen
  \bibfield  {author} {\bibinfo {author} {\bibfnamefont {L.~P.}\ \bibnamefont
  {Singer}}, \bibinfo {author} {\bibfnamefont {H.-Y.}\ \bibnamefont {Chen}},
  \bibinfo {author} {\bibfnamefont {D.~E.}\ \bibnamefont {Holz}}, \bibinfo
  {author} {\bibfnamefont {W.~M.}\ \bibnamefont {Farr}}, \bibinfo {author}
  {\bibfnamefont {L.~R.}\ \bibnamefont {Price}}, \bibinfo {author}
  {\bibfnamefont {V.}~\bibnamefont {Raymond}}, \bibinfo {author} {\bibfnamefont
  {S.~B.}\ \bibnamefont {Cenko}}, \bibinfo {author} {\bibfnamefont
  {N.}~\bibnamefont {Gehrels}}, \bibinfo {author} {\bibfnamefont
  {J.}~\bibnamefont {Cannizzo}}, \bibinfo {author} {\bibfnamefont {M.~M.}\
  \bibnamefont {Kasliwal}}, \ and\ \bibinfo {author} {\bibnamefont {et~al.}},\
  }\href {\doibase 10.3847/2041-8205/829/1/l15} {\bibfield  {journal} {\bibinfo
   {journal} {The Astrophysical Journal}\ }\textbf {\bibinfo {volume} {829}},\
  \bibinfo {pages} {L15} (\bibinfo {year} {2016})}\BibitemShut {NoStop}%
\bibitem [{\citenamefont {Chen}\ \emph {et~al.}(2021)\citenamefont {Chen},
  \citenamefont {Holz}, \citenamefont {Miller}, \citenamefont {Evans},
  \citenamefont {Vitale},\ and\ \citenamefont {Creighton}}]{sensitivity_2}%
  \BibitemOpen
  \bibfield  {author} {\bibinfo {author} {\bibfnamefont {H.-Y.}\ \bibnamefont
  {Chen}}, \bibinfo {author} {\bibfnamefont {D.~E.}\ \bibnamefont {Holz}},
  \bibinfo {author} {\bibfnamefont {J.}~\bibnamefont {Miller}}, \bibinfo
  {author} {\bibfnamefont {M.}~\bibnamefont {Evans}}, \bibinfo {author}
  {\bibfnamefont {S.}~\bibnamefont {Vitale}}, \ and\ \bibinfo {author}
  {\bibfnamefont {J.}~\bibnamefont {Creighton}},\ }\href {\doibase
  10.1088/1361-6382/abd594} {\bibfield  {journal} {\bibinfo  {journal}
  {Classical and Quantum Gravity}\ }\textbf {\bibinfo {volume} {38}},\ \bibinfo
  {pages} {055010} (\bibinfo {year} {2021})}\BibitemShut {NoStop}%
\bibitem [{\citenamefont {Ashton}\ \emph {et~al.}(2019)\citenamefont {Ashton},
  \citenamefont {Hübner}, \citenamefont {Lasky}, \citenamefont {Talbot},
  \citenamefont {Ackley}, \citenamefont {Biscoveanu}, \citenamefont {Chu},
  \citenamefont {Divakarla}, \citenamefont {Easter}, \citenamefont {Goncharov},
  \citenamefont {Vivanco}, \citenamefont {Harms}, \citenamefont {Lower},
  \citenamefont {Meadors}, \citenamefont {Melchor}, \citenamefont {Payne},
  \citenamefont {Pitkin}, \citenamefont {Powell}, \citenamefont {Sarin},
  \citenamefont {Smith},\ and\ \citenamefont {Thrane}}]{bilby_para_est}%
  \BibitemOpen
  \bibfield  {author} {\bibinfo {author} {\bibfnamefont {G.}~\bibnamefont
  {Ashton}}, \bibinfo {author} {\bibfnamefont {M.}~\bibnamefont {Hübner}},
  \bibinfo {author} {\bibfnamefont {P.~D.}\ \bibnamefont {Lasky}}, \bibinfo
  {author} {\bibfnamefont {C.}~\bibnamefont {Talbot}}, \bibinfo {author}
  {\bibfnamefont {K.}~\bibnamefont {Ackley}}, \bibinfo {author} {\bibfnamefont
  {S.}~\bibnamefont {Biscoveanu}}, \bibinfo {author} {\bibfnamefont
  {Q.}~\bibnamefont {Chu}}, \bibinfo {author} {\bibfnamefont {A.}~\bibnamefont
  {Divakarla}}, \bibinfo {author} {\bibfnamefont {P.~J.}\ \bibnamefont
  {Easter}}, \bibinfo {author} {\bibfnamefont {B.}~\bibnamefont {Goncharov}},
  \bibinfo {author} {\bibfnamefont {F.~H.}\ \bibnamefont {Vivanco}}, \bibinfo
  {author} {\bibfnamefont {J.}~\bibnamefont {Harms}}, \bibinfo {author}
  {\bibfnamefont {M.~E.}\ \bibnamefont {Lower}}, \bibinfo {author}
  {\bibfnamefont {G.~D.}\ \bibnamefont {Meadors}}, \bibinfo {author}
  {\bibfnamefont {D.}~\bibnamefont {Melchor}}, \bibinfo {author} {\bibfnamefont
  {E.}~\bibnamefont {Payne}}, \bibinfo {author} {\bibfnamefont {M.~D.}\
  \bibnamefont {Pitkin}}, \bibinfo {author} {\bibfnamefont {J.}~\bibnamefont
  {Powell}}, \bibinfo {author} {\bibfnamefont {N.}~\bibnamefont {Sarin}},
  \bibinfo {author} {\bibfnamefont {R.~J.~E.}\ \bibnamefont {Smith}}, \ and\
  \bibinfo {author} {\bibfnamefont {E.}~\bibnamefont {Thrane}},\ }\href
  {\doibase 10.3847/1538-4365/ab06fc} {\ \textbf {\bibinfo {volume} {241}},\
  \bibinfo {pages} {27} (\bibinfo {year} {2019})}\BibitemShut {NoStop}%
\bibitem [{\citenamefont {Veitch}\ \emph {et~al.}(2015)\citenamefont {Veitch},
  \citenamefont {Raymond}, \citenamefont {Farr}, \citenamefont {Farr},
  \citenamefont {Graff}, \citenamefont {Vitale}, \citenamefont {Aylott},
  \citenamefont {Blackburn}, \citenamefont {Christensen}, \citenamefont
  {Coughlin}, \citenamefont {Del~Pozzo}, \citenamefont {Feroz}, \citenamefont
  {Gair}, \citenamefont {Haster}, \citenamefont {Kalogera}, \citenamefont
  {Littenberg}, \citenamefont {Mandel}, \citenamefont {O'Shaughnessy},
  \citenamefont {Pitkin}, \citenamefont {Rodriguez}, \citenamefont {R\"over},
  \citenamefont {Sidery}, \citenamefont {Smith}, \citenamefont {Van Der~Sluys},
  \citenamefont {Vecchio}, \citenamefont {Vousden},\ and\ \citenamefont
  {Wade}}]{PhysRevD.91.042003}%
  \BibitemOpen
  \bibfield  {author} {\bibinfo {author} {\bibfnamefont {J.}~\bibnamefont
  {Veitch}}, \bibinfo {author} {\bibfnamefont {V.}~\bibnamefont {Raymond}},
  \bibinfo {author} {\bibfnamefont {B.}~\bibnamefont {Farr}}, \bibinfo {author}
  {\bibfnamefont {W.}~\bibnamefont {Farr}}, \bibinfo {author} {\bibfnamefont
  {P.}~\bibnamefont {Graff}}, \bibinfo {author} {\bibfnamefont
  {S.}~\bibnamefont {Vitale}}, \bibinfo {author} {\bibfnamefont
  {B.}~\bibnamefont {Aylott}}, \bibinfo {author} {\bibfnamefont
  {K.}~\bibnamefont {Blackburn}}, \bibinfo {author} {\bibfnamefont
  {N.}~\bibnamefont {Christensen}}, \bibinfo {author} {\bibfnamefont
  {M.}~\bibnamefont {Coughlin}}, \bibinfo {author} {\bibfnamefont
  {W.}~\bibnamefont {Del~Pozzo}}, \bibinfo {author} {\bibfnamefont
  {F.}~\bibnamefont {Feroz}}, \bibinfo {author} {\bibfnamefont
  {J.}~\bibnamefont {Gair}}, \bibinfo {author} {\bibfnamefont {C.-J.}\
  \bibnamefont {Haster}}, \bibinfo {author} {\bibfnamefont {V.}~\bibnamefont
  {Kalogera}}, \bibinfo {author} {\bibfnamefont {T.}~\bibnamefont
  {Littenberg}}, \bibinfo {author} {\bibfnamefont {I.}~\bibnamefont {Mandel}},
  \bibinfo {author} {\bibfnamefont {R.}~\bibnamefont {O'Shaughnessy}}, \bibinfo
  {author} {\bibfnamefont {M.}~\bibnamefont {Pitkin}}, \bibinfo {author}
  {\bibfnamefont {C.}~\bibnamefont {Rodriguez}}, \bibinfo {author}
  {\bibfnamefont {C.}~\bibnamefont {R\"over}}, \bibinfo {author} {\bibfnamefont
  {T.}~\bibnamefont {Sidery}}, \bibinfo {author} {\bibfnamefont
  {R.}~\bibnamefont {Smith}}, \bibinfo {author} {\bibfnamefont
  {M.}~\bibnamefont {Van Der~Sluys}}, \bibinfo {author} {\bibfnamefont
  {A.}~\bibnamefont {Vecchio}}, \bibinfo {author} {\bibfnamefont
  {W.}~\bibnamefont {Vousden}}, \ and\ \bibinfo {author} {\bibfnamefont
  {L.}~\bibnamefont {Wade}},\ }\href {\doibase 10.1103/PhysRevD.91.042003}
  {\bibfield  {journal} {\bibinfo  {journal} {Phys. Rev. D}\ }\textbf {\bibinfo
  {volume} {91}},\ \bibinfo {pages} {042003} (\bibinfo {year}
  {2015})}\BibitemShut {NoStop}%
\bibitem [{\citenamefont {Farr}(2019)}]{selectionbias}%
  \BibitemOpen
  \bibfield  {author} {\bibinfo {author} {\bibfnamefont {W.~M.}\ \bibnamefont
  {Farr}},\ }\href {\doibase 10.3847/2515-5172/ab1d5f} {\bibfield  {journal}
  {\bibinfo  {journal} {Research Notes of the AAS}\ }\textbf {\bibinfo {volume}
  {3}},\ \bibinfo {pages} {66} (\bibinfo {year} {2019})}\BibitemShut {NoStop}%
\bibitem [{\citenamefont {Gerosa}(2017)}]{davide_gerosa_2017_889966}%
  \BibitemOpen
  \bibfield  {author} {\bibinfo {author} {\bibfnamefont {D.}~\bibnamefont
  {Gerosa}},\ }\href {\doibase 10.5281/zenodo.889966} {\enquote {\bibinfo
  {title} {dgerosa/gwdet: v0.1},}\ } (\bibinfo {year} {2017})\BibitemShut
  {NoStop}%
\bibitem [{\citenamefont {Belczynski}\ \emph {et~al.}(2008)\citenamefont
  {Belczynski}, \citenamefont {Kalogera}, \citenamefont {Rasio}, \citenamefont
  {Taam}, \citenamefont {Zezas}, \citenamefont {Bulik}, \citenamefont
  {Maccarone},\ and\ \citenamefont {Ivanova}}]{popsyn_compactObjectModeling}%
  \BibitemOpen
  \bibfield  {author} {\bibinfo {author} {\bibfnamefont {K.}~\bibnamefont
  {Belczynski}}, \bibinfo {author} {\bibfnamefont {V.}~\bibnamefont
  {Kalogera}}, \bibinfo {author} {\bibfnamefont {F.~A.}\ \bibnamefont {Rasio}},
  \bibinfo {author} {\bibfnamefont {R.~E.}\ \bibnamefont {Taam}}, \bibinfo
  {author} {\bibfnamefont {A.}~\bibnamefont {Zezas}}, \bibinfo {author}
  {\bibfnamefont {T.}~\bibnamefont {Bulik}}, \bibinfo {author} {\bibfnamefont
  {T.~J.}\ \bibnamefont {Maccarone}}, \ and\ \bibinfo {author} {\bibfnamefont
  {N.}~\bibnamefont {Ivanova}},\ }\href {\doibase 10.1086/521026} {\bibfield
  {journal} {\bibinfo  {journal} {The Astrophysical Journal Supplement Series}\
  }\textbf {\bibinfo {volume} {174}},\ \bibinfo {pages} {223–260} (\bibinfo
  {year} {2008})}\BibitemShut {NoStop}%
\bibitem [{\citenamefont {Hurley}\ \emph {et~al.}(2002)\citenamefont {Hurley},
  \citenamefont {Tout},\ and\ \citenamefont {Pols}}]{popsyn_EvolutionOfBinary}%
  \BibitemOpen
  \bibfield  {author} {\bibinfo {author} {\bibfnamefont {J.~R.}\ \bibnamefont
  {Hurley}}, \bibinfo {author} {\bibfnamefont {C.~A.}\ \bibnamefont {Tout}}, \
  and\ \bibinfo {author} {\bibfnamefont {O.~R.}\ \bibnamefont {Pols}},\ }\href
  {\doibase 10.1046/j.1365-8711.2002.05038.x} {\bibfield  {journal} {\bibinfo
  {journal} {Monthly Notices of the Royal Astronomical Society}\ }\textbf
  {\bibinfo {volume} {329}},\ \bibinfo {pages} {897} (\bibinfo {year}
  {2002})},\ \Eprint
  {http://arxiv.org/abs/https://academic.oup.com/mnras/article-pdf/329/4/897/18418535/329-4-897.pdf}
  {https://academic.oup.com/mnras/article-pdf/329/4/897/18418535/329-4-897.pdf}
  \BibitemShut {NoStop}%
\bibitem [{\citenamefont {Giersz}\ \emph {et~al.}(2013)\citenamefont {Giersz},
  \citenamefont {Heggie}, \citenamefont {Hurley},\ and\ \citenamefont
  {Hypki}}]{popsyn_MOCCA}%
  \BibitemOpen
  \bibfield  {author} {\bibinfo {author} {\bibfnamefont {M.}~\bibnamefont
  {Giersz}}, \bibinfo {author} {\bibfnamefont {D.~C.}\ \bibnamefont {Heggie}},
  \bibinfo {author} {\bibfnamefont {J.~R.}\ \bibnamefont {Hurley}}, \ and\
  \bibinfo {author} {\bibfnamefont {A.}~\bibnamefont {Hypki}},\ }\href
  {\doibase 10.1093/mnras/stt307} {\bibfield  {journal} {\bibinfo  {journal}
  {Monthly Notices of the Royal Astronomical Society}\ }\textbf {\bibinfo
  {volume} {431}},\ \bibinfo {pages} {2184} (\bibinfo {year} {2013})},\ \Eprint
  {http://arxiv.org/abs/https://academic.oup.com/mnras/article-pdf/431/3/2184/4895448/stt307.pdf}
  {https://academic.oup.com/mnras/article-pdf/431/3/2184/4895448/stt307.pdf}
  \BibitemShut {NoStop}%
\bibitem [{\citenamefont {Giacobbo}\ and\ \citenamefont
  {Mapelli}(2018)}]{popsyn_progenitorsOfCompactObject}%
  \BibitemOpen
  \bibfield  {author} {\bibinfo {author} {\bibfnamefont {N.}~\bibnamefont
  {Giacobbo}}\ and\ \bibinfo {author} {\bibfnamefont {M.}~\bibnamefont
  {Mapelli}},\ }\href {\doibase 10.1093/mnras/sty1999} {\bibfield  {journal}
  {\bibinfo  {journal} {Monthly Notices of the Royal Astronomical Society}\
  }\textbf {\bibinfo {volume} {480}},\ \bibinfo {pages} {2011} (\bibinfo {year}
  {2018})},\ \Eprint
  {http://arxiv.org/abs/https://academic.oup.com/mnras/article-pdf/480/2/2011/25440572/sty1999.pdf}
  {https://academic.oup.com/mnras/article-pdf/480/2/2011/25440572/sty1999.pdf}
  \BibitemShut {NoStop}%
\bibitem [{\citenamefont {Donovan}\ \emph {et~al.}(2015)\citenamefont
  {Donovan}, \citenamefont {Burrage}, \citenamefont {Burrage}, \citenamefont
  {McCourt}, \citenamefont {Thompson},\ and\ \citenamefont
  {Yazici}}]{lhsestimates}%
  \BibitemOpen
  \bibfield  {author} {\bibinfo {author} {\bibfnamefont {D.}~\bibnamefont
  {Donovan}}, \bibinfo {author} {\bibfnamefont {K.}~\bibnamefont {Burrage}},
  \bibinfo {author} {\bibfnamefont {P.}~\bibnamefont {Burrage}}, \bibinfo
  {author} {\bibfnamefont {T.~A.}\ \bibnamefont {McCourt}}, \bibinfo {author}
  {\bibfnamefont {H.~B.}\ \bibnamefont {Thompson}}, \ and\ \bibinfo {author}
  {\bibfnamefont {E.~S.}\ \bibnamefont {Yazici}},\ }\href@noop {} {\enquote
  {\bibinfo {title} {Estimates of the coverage of parameter space by latin
  hypercube and orthogonal sampling: connections between populations of models
  and experimental designs},}\ } (\bibinfo {year} {2015}),\ \Eprint
  {http://arxiv.org/abs/1510.03502} {arXiv:1510.03502 [math.ST]} \BibitemShut
  {NoStop}%
\bibitem [{\citenamefont {Stein}(1987)}]{lhs_stein}%
  \BibitemOpen
  \bibfield  {author} {\bibinfo {author} {\bibfnamefont {M.}~\bibnamefont
  {Stein}},\ }\href {\doibase 10.1080/00401706.1987.10488205} {\bibfield
  {journal} {\bibinfo  {journal} {Technometrics}\ }\textbf {\bibinfo {volume}
  {29}},\ \bibinfo {pages} {143} (\bibinfo {year} {1987})}\BibitemShut
  {NoStop}%
\bibitem [{\citenamefont {Lee}(2014)}]{pyDOE}%
  \BibitemOpen
  \bibfield  {author} {\bibinfo {author} {\bibfnamefont {A.~D.}\ \bibnamefont
  {Lee}},\ }\href {https://github.com/tisimst/pyDOE} {\enquote {\bibinfo
  {title} {pydoe,the experimental design package for python},}\ } (\bibinfo
  {year} {2014})\BibitemShut {NoStop}%
\bibitem [{\citenamefont {Rasmussen}\ \emph {et~al.}(2016)\citenamefont
  {Rasmussen}, \citenamefont {E.}, ,\ and\ \citenamefont {Williams}}]{gpr}%
  \BibitemOpen
  \bibfield  {author} {\bibinfo {author} {\bibnamefont {Rasmussen}}, \bibinfo
  {author} {\bibfnamefont {C.}~\bibnamefont {E.}}, , \ and\ \bibinfo {author}
  {\bibfnamefont {C.~K.~I.}\ \bibnamefont {Williams}},\ }\href@noop {} {\
  (\bibinfo {year} {2016})}\BibitemShut {NoStop}%
\bibitem [{\citenamefont {Buitinck}\ \emph {et~al.}(2013)\citenamefont
  {Buitinck}, \citenamefont {Louppe}, \citenamefont {Blondel}, \citenamefont
  {Pedregosa}, \citenamefont {Mueller}, \citenamefont {Grisel}, \citenamefont
  {Niculae}, \citenamefont {Prettenhofer}, \citenamefont {Gramfort},
  \citenamefont {Grobler}, \citenamefont {Layton}, \citenamefont {Vanderplas},
  \citenamefont {Joly}, \citenamefont {Holt},\ and\ \citenamefont
  {Varoquaux}}]{scikitlearn}%
  \BibitemOpen
  \bibfield  {author} {\bibinfo {author} {\bibfnamefont {L.}~\bibnamefont
  {Buitinck}}, \bibinfo {author} {\bibfnamefont {G.}~\bibnamefont {Louppe}},
  \bibinfo {author} {\bibfnamefont {M.}~\bibnamefont {Blondel}}, \bibinfo
  {author} {\bibfnamefont {F.}~\bibnamefont {Pedregosa}}, \bibinfo {author}
  {\bibfnamefont {A.}~\bibnamefont {Mueller}}, \bibinfo {author} {\bibfnamefont
  {O.}~\bibnamefont {Grisel}}, \bibinfo {author} {\bibfnamefont
  {V.}~\bibnamefont {Niculae}}, \bibinfo {author} {\bibfnamefont
  {P.}~\bibnamefont {Prettenhofer}}, \bibinfo {author} {\bibfnamefont
  {A.}~\bibnamefont {Gramfort}}, \bibinfo {author} {\bibfnamefont
  {J.}~\bibnamefont {Grobler}}, \bibinfo {author} {\bibfnamefont
  {R.}~\bibnamefont {Layton}}, \bibinfo {author} {\bibfnamefont
  {J.}~\bibnamefont {Vanderplas}}, \bibinfo {author} {\bibfnamefont
  {A.}~\bibnamefont {Joly}}, \bibinfo {author} {\bibfnamefont {B.}~\bibnamefont
  {Holt}}, \ and\ \bibinfo {author} {\bibfnamefont {G.}~\bibnamefont
  {Varoquaux}},\ }\href@noop {} {\enquote {\bibinfo {title} {Api design for
  machine learning software: experiences from the scikit-learn project},}\ }
  (\bibinfo {year} {2013}),\ \Eprint {http://arxiv.org/abs/1309.0238}
  {arXiv:1309.0238 [cs.LG]} \BibitemShut {NoStop}%
\bibitem [{\citenamefont {Papamakarios}\ \emph {et~al.}(2018)\citenamefont
  {Papamakarios}, \citenamefont {Pavlakou},\ and\ \citenamefont
  {Murray}}]{MAF}%
  \BibitemOpen
  \bibfield  {author} {\bibinfo {author} {\bibfnamefont {G.}~\bibnamefont
  {Papamakarios}}, \bibinfo {author} {\bibfnamefont {T.}~\bibnamefont
  {Pavlakou}}, \ and\ \bibinfo {author} {\bibfnamefont {I.}~\bibnamefont
  {Murray}},\ }\href@noop {} {\enquote {\bibinfo {title} {Masked autoregressive
  flow for density estimation},}\ } (\bibinfo {year} {2018}),\ \Eprint
  {http://arxiv.org/abs/1705.07057} {arXiv:1705.07057 [stat.ML]} \BibitemShut
  {NoStop}%
\bibitem [{\citenamefont {Kingma}\ and\ \citenamefont
  {Welling}(2014)}]{variationalBayes}%
  \BibitemOpen
  \bibfield  {author} {\bibinfo {author} {\bibfnamefont {D.~P.}\ \bibnamefont
  {Kingma}}\ and\ \bibinfo {author} {\bibfnamefont {M.}~\bibnamefont
  {Welling}},\ }\href@noop {} {\enquote {\bibinfo {title} {Auto-encoding
  variational bayes},}\ } (\bibinfo {year} {2014}),\ \Eprint
  {http://arxiv.org/abs/1312.6114} {arXiv:1312.6114 [stat.ML]} \BibitemShut
  {NoStop}%
\bibitem [{\citenamefont {Goodfellow}\ \emph {et~al.}(2014)\citenamefont
  {Goodfellow}, \citenamefont {Pouget-Abadie}, \citenamefont {Mirza},
  \citenamefont {Xu}, \citenamefont {Warde-Farley}, \citenamefont {Ozair},
  \citenamefont {Courville},\ and\ \citenamefont {Bengio}}]{adversarial}%
  \BibitemOpen
  \bibfield  {author} {\bibinfo {author} {\bibfnamefont {I.~J.}\ \bibnamefont
  {Goodfellow}}, \bibinfo {author} {\bibfnamefont {J.}~\bibnamefont
  {Pouget-Abadie}}, \bibinfo {author} {\bibfnamefont {M.}~\bibnamefont
  {Mirza}}, \bibinfo {author} {\bibfnamefont {B.}~\bibnamefont {Xu}}, \bibinfo
  {author} {\bibfnamefont {D.}~\bibnamefont {Warde-Farley}}, \bibinfo {author}
  {\bibfnamefont {S.}~\bibnamefont {Ozair}}, \bibinfo {author} {\bibfnamefont
  {A.}~\bibnamefont {Courville}}, \ and\ \bibinfo {author} {\bibfnamefont
  {Y.}~\bibnamefont {Bengio}},\ }\href@noop {} {\enquote {\bibinfo {title}
  {Generative adversarial networks},}\ } (\bibinfo {year} {2014}),\ \Eprint
  {http://arxiv.org/abs/1406.2661} {arXiv:1406.2661 [stat.ML]} \BibitemShut
  {NoStop}%
\bibitem [{\citenamefont {Papamakarios}\ \emph {et~al.}(2017)\citenamefont
  {Papamakarios}, \citenamefont {Pavlakou},\ and\ \citenamefont
  {Murray}}]{papamakarios2017masked}%
  \BibitemOpen
  \bibfield  {author} {\bibinfo {author} {\bibfnamefont {G.}~\bibnamefont
  {Papamakarios}}, \bibinfo {author} {\bibfnamefont {T.}~\bibnamefont
  {Pavlakou}}, \ and\ \bibinfo {author} {\bibfnamefont {I.}~\bibnamefont
  {Murray}},\ }\href@noop {} {\bibfield  {journal} {\bibinfo  {journal} {arXiv
  preprint arXiv:1705.07057}\ } (\bibinfo {year} {2017})}\BibitemShut {NoStop}%
\bibitem [{\citenamefont {Uria}\ \emph {et~al.}(2016)\citenamefont {Uria},
  \citenamefont {Côté}, \citenamefont {Gregor}, \citenamefont {Murray},\ and\
  \citenamefont {Larochelle}}]{MADE_product}%
  \BibitemOpen
  \bibfield  {author} {\bibinfo {author} {\bibfnamefont {B.}~\bibnamefont
  {Uria}}, \bibinfo {author} {\bibfnamefont {M.-A.}\ \bibnamefont {Côté}},
  \bibinfo {author} {\bibfnamefont {K.}~\bibnamefont {Gregor}}, \bibinfo
  {author} {\bibfnamefont {I.}~\bibnamefont {Murray}}, \ and\ \bibinfo {author}
  {\bibfnamefont {H.}~\bibnamefont {Larochelle}},\ }\href@noop {} {\enquote
  {\bibinfo {title} {Neural autoregressive distribution estimation},}\ }
  (\bibinfo {year} {2016}),\ \Eprint {http://arxiv.org/abs/1605.02226}
  {arXiv:1605.02226 [cs.LG]} \BibitemShut {NoStop}%
\bibitem [{\citenamefont {Liao}\ and\ \citenamefont
  {He}(2021)}]{liao2021jacobian}%
  \BibitemOpen
  \bibfield  {author} {\bibinfo {author} {\bibfnamefont {H.}~\bibnamefont
  {Liao}}\ and\ \bibinfo {author} {\bibfnamefont {J.}~\bibnamefont {He}},\
  }\href@noop {} {\enquote {\bibinfo {title} {Jacobian determinant of
  normalizing flows},}\ } (\bibinfo {year} {2021}),\ \Eprint
  {http://arxiv.org/abs/2102.06539} {arXiv:2102.06539 [cs.LG]} \BibitemShut
  {NoStop}%
\bibitem [{\citenamefont {Kullback}\ and\ \citenamefont
  {Leibler}(1951)}]{KLdiv}%
  \BibitemOpen
  \bibfield  {author} {\bibinfo {author} {\bibfnamefont {S.}~\bibnamefont
  {Kullback}}\ and\ \bibinfo {author} {\bibfnamefont {R.~A.}\ \bibnamefont
  {Leibler}},\ }\href@noop {} {\bibfield  {journal} {\bibinfo  {journal} {The
  annals of mathematical statistics}\ }\textbf {\bibinfo {volume} {22}},\
  \bibinfo {pages} {79} (\bibinfo {year} {1951})}\BibitemShut {NoStop}%
\bibitem [{\citenamefont {Liu}\ \emph {et~al.}(2019)\citenamefont {Liu},
  \citenamefont {Ong}, \citenamefont {Shen},\ and\ \citenamefont
  {Cai}}]{liugprlimitation}%
  \BibitemOpen
  \bibfield  {author} {\bibinfo {author} {\bibfnamefont {H.}~\bibnamefont
  {Liu}}, \bibinfo {author} {\bibfnamefont {Y.-S.}\ \bibnamefont {Ong}},
  \bibinfo {author} {\bibfnamefont {X.}~\bibnamefont {Shen}}, \ and\ \bibinfo
  {author} {\bibfnamefont {J.}~\bibnamefont {Cai}},\ }\href@noop {} {\enquote
  {\bibinfo {title} {When gaussian process meets big data: A review of scalable
  gps},}\ } (\bibinfo {year} {2019}),\ \Eprint
  {http://arxiv.org/abs/1807.01065} {arXiv:1807.01065 [stat.ML]} \BibitemShut
  {NoStop}%
\bibitem [{\citenamefont {Swiler}\ \emph {et~al.}(2020)\citenamefont {Swiler},
  \citenamefont {Gulian}, \citenamefont {Frankel}, \citenamefont {Safta},\ and\
  \citenamefont {Jakeman}}]{2020gprlimitation}%
  \BibitemOpen
  \bibfield  {author} {\bibinfo {author} {\bibfnamefont {L.~P.}\ \bibnamefont
  {Swiler}}, \bibinfo {author} {\bibfnamefont {M.}~\bibnamefont {Gulian}},
  \bibinfo {author} {\bibfnamefont {A.~L.}\ \bibnamefont {Frankel}}, \bibinfo
  {author} {\bibfnamefont {C.}~\bibnamefont {Safta}}, \ and\ \bibinfo {author}
  {\bibfnamefont {J.~D.}\ \bibnamefont {Jakeman}},\ }\href {\doibase
  10.1615/jmachlearnmodelcomput.2020035155} {\bibfield  {journal} {\bibinfo
  {journal} {Journal of Machine Learning for Modeling and Computing}\ }\textbf
  {\bibinfo {volume} {1}},\ \bibinfo {pages} {119–156} (\bibinfo {year}
  {2020})}\BibitemShut {NoStop}%
\bibitem [{\citenamefont {Hermans}\ \emph {et~al.}(2021)\citenamefont
  {Hermans}, \citenamefont {Delaunoy}, \citenamefont {Rozet}, \citenamefont
  {Wehenkel},\ and\ \citenamefont {Louppe}}]{SBI_crisis}%
  \BibitemOpen
  \bibfield  {author} {\bibinfo {author} {\bibfnamefont {J.}~\bibnamefont
  {Hermans}}, \bibinfo {author} {\bibfnamefont {A.}~\bibnamefont {Delaunoy}},
  \bibinfo {author} {\bibfnamefont {F.}~\bibnamefont {Rozet}}, \bibinfo
  {author} {\bibfnamefont {A.}~\bibnamefont {Wehenkel}}, \ and\ \bibinfo
  {author} {\bibfnamefont {G.}~\bibnamefont {Louppe}},\ }\href {\doibase
  10.48550/ARXIV.2110.06581} {\enquote {\bibinfo {title} {Averting a crisis in
  simulation-based inference},}\ } (\bibinfo {year} {2021})\BibitemShut
  {NoStop}%
\bibitem [{\citenamefont {Giacobbo}\ \emph {et~al.}(2018)\citenamefont
  {Giacobbo}, \citenamefont {Mapelli},\ and\ \citenamefont {Spera}}]{m2}%
  \BibitemOpen
  \bibfield  {author} {\bibinfo {author} {\bibfnamefont {N.}~\bibnamefont
  {Giacobbo}}, \bibinfo {author} {\bibfnamefont {M.}~\bibnamefont {Mapelli}}, \
  and\ \bibinfo {author} {\bibfnamefont {M.}~\bibnamefont {Spera}},\
  }\href@noop {} {\bibfield  {journal} {\bibinfo  {journal} {Monthly Notices of
  the Royal Astronomical Society}\ }\textbf {\bibinfo {volume} {474}},\
  \bibinfo {pages} {2959} (\bibinfo {year} {2018})}\BibitemShut {NoStop}%
\bibitem [{\citenamefont {Breivik}\ \emph {et~al.}(2020)\citenamefont
  {Breivik}, \citenamefont {Coughlin}, \citenamefont {Zevin}, \citenamefont
  {Rodriguez}, \citenamefont {Kremer}, \citenamefont {Ye}, \citenamefont
  {Andrews}, \citenamefont {Kurkowski}, \citenamefont {Digman}, \citenamefont
  {Larson},\ and\ \citenamefont {et~al.}}]{m3}%
  \BibitemOpen
  \bibfield  {author} {\bibinfo {author} {\bibfnamefont {K.}~\bibnamefont
  {Breivik}}, \bibinfo {author} {\bibfnamefont {S.}~\bibnamefont {Coughlin}},
  \bibinfo {author} {\bibfnamefont {M.}~\bibnamefont {Zevin}}, \bibinfo
  {author} {\bibfnamefont {C.~L.}\ \bibnamefont {Rodriguez}}, \bibinfo {author}
  {\bibfnamefont {K.}~\bibnamefont {Kremer}}, \bibinfo {author} {\bibfnamefont
  {C.~S.}\ \bibnamefont {Ye}}, \bibinfo {author} {\bibfnamefont {J.~J.}\
  \bibnamefont {Andrews}}, \bibinfo {author} {\bibfnamefont {M.}~\bibnamefont
  {Kurkowski}}, \bibinfo {author} {\bibfnamefont {M.~C.}\ \bibnamefont
  {Digman}}, \bibinfo {author} {\bibfnamefont {S.~L.}\ \bibnamefont {Larson}},
  \ and\ \bibinfo {author} {\bibnamefont {et~al.}},\ }\href {\doibase
  10.3847/1538-4357/ab9d85} {\bibfield  {journal} {\bibinfo  {journal} {The
  Astrophysical Journal}\ }\textbf {\bibinfo {volume} {898}},\ \bibinfo {pages}
  {71} (\bibinfo {year} {2020})}\BibitemShut {NoStop}%
\bibitem [{\citenamefont {Dominik}\ \emph {et~al.}(2015)\citenamefont
  {Dominik}, \citenamefont {Berti}, \citenamefont {O'Shaughnessy},
  \citenamefont {Mandel}, \citenamefont {Belczynski}, \citenamefont {Fryer},
  \citenamefont {Holz}, \citenamefont {Bulik},\ and\ \citenamefont
  {Pannarale}}]{m4}%
  \BibitemOpen
  \bibfield  {author} {\bibinfo {author} {\bibfnamefont {M.}~\bibnamefont
  {Dominik}}, \bibinfo {author} {\bibfnamefont {E.}~\bibnamefont {Berti}},
  \bibinfo {author} {\bibfnamefont {R.}~\bibnamefont {O'Shaughnessy}}, \bibinfo
  {author} {\bibfnamefont {I.}~\bibnamefont {Mandel}}, \bibinfo {author}
  {\bibfnamefont {K.}~\bibnamefont {Belczynski}}, \bibinfo {author}
  {\bibfnamefont {C.}~\bibnamefont {Fryer}}, \bibinfo {author} {\bibfnamefont
  {D.~E.}\ \bibnamefont {Holz}}, \bibinfo {author} {\bibfnamefont
  {T.}~\bibnamefont {Bulik}}, \ and\ \bibinfo {author} {\bibfnamefont
  {F.}~\bibnamefont {Pannarale}},\ }\href {\doibase
  10.1088/0004-637x/806/2/263} {\ \textbf {\bibinfo {volume} {806}},\ \bibinfo
  {pages} {263} (\bibinfo {year} {2015})}\BibitemShut {NoStop}%
\bibitem [{\citenamefont {Jenkins}\ and\ \citenamefont
  {Peacock}(2011)}]{model_selection_bayes}%
  \BibitemOpen
  \bibfield  {author} {\bibinfo {author} {\bibfnamefont {C.~R.}\ \bibnamefont
  {Jenkins}}\ and\ \bibinfo {author} {\bibfnamefont {J.~A.}\ \bibnamefont
  {Peacock}},\ }\href {\doibase 10.1111/j.1365-2966.2011.18361.x} {\bibfield
  {journal} {\bibinfo  {journal} {Monthly Notices of the Royal Astronomical
  Society}\ }\textbf {\bibinfo {volume} {413}},\ \bibinfo {pages} {2895}
  (\bibinfo {year} {2011})},\ \Eprint
  {http://arxiv.org/abs/https://academic.oup.com/mnras/article-pdf/413/4/2895/2882506/mnras0413-2895.pdf}
  {https://academic.oup.com/mnras/article-pdf/413/4/2895/2882506/mnras0413-2895.pdf}
  \BibitemShut {NoStop}%
\bibitem [{\citenamefont {Louizos}\ and\ \citenamefont
  {Welling}(2017)}]{NF_BNN}%
  \BibitemOpen
  \bibfield  {author} {\bibinfo {author} {\bibfnamefont {C.}~\bibnamefont
  {Louizos}}\ and\ \bibinfo {author} {\bibfnamefont {M.}~\bibnamefont
  {Welling}},\ }\href@noop {} {\enquote {\bibinfo {title} {Multiplicative
  normalizing flows for variational bayesian neural networks},}\ } (\bibinfo
  {year} {2017}),\ \Eprint {http://arxiv.org/abs/1703.01961} {arXiv:1703.01961
  [stat.ML]} \BibitemShut {NoStop}%
\bibitem [{\citenamefont {Hunter}(2007)}]{matplotlib}%
  \BibitemOpen
  \bibfield  {author} {\bibinfo {author} {\bibfnamefont {J.~D.}\ \bibnamefont
  {Hunter}},\ }\href {\doibase 10.1109/MCSE.2007.55} {\bibfield  {journal}
  {\bibinfo  {journal} {Computing in Science Engineering}\ }\textbf {\bibinfo
  {volume} {9}},\ \bibinfo {pages} {90} (\bibinfo {year} {2007})}\BibitemShut
  {NoStop}%
\bibitem [{\citenamefont {van~der Walt}\ \emph {et~al.}(2011)\citenamefont
  {van~der Walt}, \citenamefont {Colbert},\ and\ \citenamefont
  {Varoquaux}}]{numpy}%
  \BibitemOpen
  \bibfield  {author} {\bibinfo {author} {\bibfnamefont {S.}~\bibnamefont
  {van~der Walt}}, \bibinfo {author} {\bibfnamefont {S.~C.}\ \bibnamefont
  {Colbert}}, \ and\ \bibinfo {author} {\bibfnamefont {G.}~\bibnamefont
  {Varoquaux}},\ }\href {\doibase 10.1109/MCSE.2011.37} {\bibfield  {journal}
  {\bibinfo  {journal} {Computing in Science Engineering}\ }\textbf {\bibinfo
  {volume} {13}},\ \bibinfo {pages} {22} (\bibinfo {year} {2011})}\BibitemShut
  {NoStop}%
\bibitem [{\citenamefont {Virtanen}\ \emph {et~al.}(2020)\citenamefont
  {Virtanen}, \citenamefont {Gommers}, \citenamefont {Oliphant}, \citenamefont
  {Haberland}, \citenamefont {Reddy}, \citenamefont {Cournapeau}, \citenamefont
  {Burovski}, \citenamefont {Peterson}, \citenamefont {Weckesser},
  \citenamefont {Bright}, \citenamefont {{van der Walt}}, \citenamefont
  {Brett}, \citenamefont {Wilson}, \citenamefont {Millman}, \citenamefont
  {Mayorov}, \citenamefont {Nelson}, \citenamefont {Jones}, \citenamefont
  {Kern}, \citenamefont {Larson}, \citenamefont {Carey}, \citenamefont {Polat},
  \citenamefont {Feng}, \citenamefont {Moore}, \citenamefont {{VanderPlas}},
  \citenamefont {Laxalde}, \citenamefont {Perktold}, \citenamefont {Cimrman},
  \citenamefont {Henriksen}, \citenamefont {Quintero}, \citenamefont {Harris},
  \citenamefont {Archibald}, \citenamefont {Ribeiro}, \citenamefont
  {Pedregosa}, \citenamefont {{van Mulbregt}},\ and\ \citenamefont {{SciPy 1.0
  Contributors}}}]{scipy}%
  \BibitemOpen
  \bibfield  {author} {\bibinfo {author} {\bibfnamefont {P.}~\bibnamefont
  {Virtanen}}, \bibinfo {author} {\bibfnamefont {R.}~\bibnamefont {Gommers}},
  \bibinfo {author} {\bibfnamefont {T.~E.}\ \bibnamefont {Oliphant}}, \bibinfo
  {author} {\bibfnamefont {M.}~\bibnamefont {Haberland}}, \bibinfo {author}
  {\bibfnamefont {T.}~\bibnamefont {Reddy}}, \bibinfo {author} {\bibfnamefont
  {D.}~\bibnamefont {Cournapeau}}, \bibinfo {author} {\bibfnamefont
  {E.}~\bibnamefont {Burovski}}, \bibinfo {author} {\bibfnamefont
  {P.}~\bibnamefont {Peterson}}, \bibinfo {author} {\bibfnamefont
  {W.}~\bibnamefont {Weckesser}}, \bibinfo {author} {\bibfnamefont
  {J.}~\bibnamefont {Bright}}, \bibinfo {author} {\bibfnamefont {S.~J.}\
  \bibnamefont {{van der Walt}}}, \bibinfo {author} {\bibfnamefont
  {M.}~\bibnamefont {Brett}}, \bibinfo {author} {\bibfnamefont
  {J.}~\bibnamefont {Wilson}}, \bibinfo {author} {\bibfnamefont {K.~J.}\
  \bibnamefont {Millman}}, \bibinfo {author} {\bibfnamefont {N.}~\bibnamefont
  {Mayorov}}, \bibinfo {author} {\bibfnamefont {A.~R.~J.}\ \bibnamefont
  {Nelson}}, \bibinfo {author} {\bibfnamefont {E.}~\bibnamefont {Jones}},
  \bibinfo {author} {\bibfnamefont {R.}~\bibnamefont {Kern}}, \bibinfo {author}
  {\bibfnamefont {E.}~\bibnamefont {Larson}}, \bibinfo {author} {\bibfnamefont
  {C.~J.}\ \bibnamefont {Carey}}, \bibinfo {author} {\bibfnamefont
  {{\.I}.}~\bibnamefont {Polat}}, \bibinfo {author} {\bibfnamefont
  {Y.}~\bibnamefont {Feng}}, \bibinfo {author} {\bibfnamefont {E.~W.}\
  \bibnamefont {Moore}}, \bibinfo {author} {\bibfnamefont {J.}~\bibnamefont
  {{VanderPlas}}}, \bibinfo {author} {\bibfnamefont {D.}~\bibnamefont
  {Laxalde}}, \bibinfo {author} {\bibfnamefont {J.}~\bibnamefont {Perktold}},
  \bibinfo {author} {\bibfnamefont {R.}~\bibnamefont {Cimrman}}, \bibinfo
  {author} {\bibfnamefont {I.}~\bibnamefont {Henriksen}}, \bibinfo {author}
  {\bibfnamefont {E.~A.}\ \bibnamefont {Quintero}}, \bibinfo {author}
  {\bibfnamefont {C.~R.}\ \bibnamefont {Harris}}, \bibinfo {author}
  {\bibfnamefont {A.~M.}\ \bibnamefont {Archibald}}, \bibinfo {author}
  {\bibfnamefont {A.~H.}\ \bibnamefont {Ribeiro}}, \bibinfo {author}
  {\bibfnamefont {F.}~\bibnamefont {Pedregosa}}, \bibinfo {author}
  {\bibfnamefont {P.}~\bibnamefont {{van Mulbregt}}}, \ and\ \bibinfo {author}
  {\bibnamefont {{SciPy 1.0 Contributors}}},\ }\href {\doibase
  10.1038/s41592-019-0686-2} {\bibfield  {journal} {\bibinfo  {journal} {Nature
  Methods}\ }\textbf {\bibinfo {volume} {17}},\ \bibinfo {pages} {261}
  (\bibinfo {year} {2020})}\BibitemShut {NoStop}%
\bibitem [{\citenamefont {Foreman-Mackey}(2016)}]{corner}%
  \BibitemOpen
  \bibfield  {author} {\bibinfo {author} {\bibfnamefont {D.}~\bibnamefont
  {Foreman-Mackey}},\ }\href {\doibase 10.21105/joss.00024} {\bibfield
  {journal} {\bibinfo  {journal} {The Journal of Open Source Software}\
  }\textbf {\bibinfo {volume} {1}},\ \bibinfo {pages} {24} (\bibinfo {year}
  {2016})}\BibitemShut {NoStop}%
\bibitem [{\citenamefont {Paszke}\ \emph {et~al.}(2019)\citenamefont {Paszke},
  \citenamefont {Gross}, \citenamefont {Massa}, \citenamefont {Lerer},
  \citenamefont {Bradbury}, \citenamefont {Chanan}, \citenamefont {Killeen},
  \citenamefont {Lin}, \citenamefont {Gimelshein}, \citenamefont {Antiga},
  \citenamefont {Desmaison}, \citenamefont {Kopf}, \citenamefont {Yang},
  \citenamefont {DeVito}, \citenamefont {Raison}, \citenamefont {Tejani},
  \citenamefont {Chilamkurthy}, \citenamefont {Steiner}, \citenamefont {Fang},
  \citenamefont {Bai},\ and\ \citenamefont {Chintala}}]{torch}%
  \BibitemOpen
  \bibfield  {author} {\bibinfo {author} {\bibfnamefont {A.}~\bibnamefont
  {Paszke}}, \bibinfo {author} {\bibfnamefont {S.}~\bibnamefont {Gross}},
  \bibinfo {author} {\bibfnamefont {F.}~\bibnamefont {Massa}}, \bibinfo
  {author} {\bibfnamefont {A.}~\bibnamefont {Lerer}}, \bibinfo {author}
  {\bibfnamefont {J.}~\bibnamefont {Bradbury}}, \bibinfo {author}
  {\bibfnamefont {G.}~\bibnamefont {Chanan}}, \bibinfo {author} {\bibfnamefont
  {T.}~\bibnamefont {Killeen}}, \bibinfo {author} {\bibfnamefont
  {Z.}~\bibnamefont {Lin}}, \bibinfo {author} {\bibfnamefont {N.}~\bibnamefont
  {Gimelshein}}, \bibinfo {author} {\bibfnamefont {L.}~\bibnamefont {Antiga}},
  \bibinfo {author} {\bibfnamefont {A.}~\bibnamefont {Desmaison}}, \bibinfo
  {author} {\bibfnamefont {A.}~\bibnamefont {Kopf}}, \bibinfo {author}
  {\bibfnamefont {E.}~\bibnamefont {Yang}}, \bibinfo {author} {\bibfnamefont
  {Z.}~\bibnamefont {DeVito}}, \bibinfo {author} {\bibfnamefont
  {M.}~\bibnamefont {Raison}}, \bibinfo {author} {\bibfnamefont
  {A.}~\bibnamefont {Tejani}}, \bibinfo {author} {\bibfnamefont
  {S.}~\bibnamefont {Chilamkurthy}}, \bibinfo {author} {\bibfnamefont
  {B.}~\bibnamefont {Steiner}}, \bibinfo {author} {\bibfnamefont
  {L.}~\bibnamefont {Fang}}, \bibinfo {author} {\bibfnamefont {J.}~\bibnamefont
  {Bai}}, \ and\ \bibinfo {author} {\bibfnamefont {S.}~\bibnamefont
  {Chintala}},\ }in\ \href
  {http://papers.neurips.cc/paper/9015-pytorch-an-imperative-style-high-performance-deep-learning-library.pdf}
  {\emph {\bibinfo {booktitle} {Advances in Neural Information Processing
  Systems 32}}},\ \bibinfo {editor} {edited by\ \bibinfo {editor}
  {\bibfnamefont {H.}~\bibnamefont {Wallach}}, \bibinfo {editor} {\bibfnamefont
  {H.}~\bibnamefont {Larochelle}}, \bibinfo {editor} {\bibfnamefont
  {A.}~\bibnamefont {Beygelzimer}}, \bibinfo {editor} {\bibfnamefont
  {F.}~\bibnamefont {d\textquotesingle Alch\'{e}-Buc}}, \bibinfo {editor}
  {\bibfnamefont {E.}~\bibnamefont {Fox}}, \ and\ \bibinfo {editor}
  {\bibfnamefont {R.}~\bibnamefont {Garnett}}}\ (\bibinfo  {publisher} {Curran
  Associates, Inc.},\ \bibinfo {year} {2019})\ pp.\ \bibinfo {pages}
  {8024--8035}\BibitemShut {NoStop}%
\end{thebibliography}%
\end{document}